\documentclass[pre, reprint,amsmath,amssymb,superscriptaddress]{revtex4-2}

\usepackage{amsmath}
\usepackage{MnSymbol}
\usepackage{dsfont}
\usepackage{graphicx} 
\usepackage{xcolor}
\usepackage[colorlinks=true,allcolors=blue]{hyperref}

\begin{document}

\title{Current fluctuations in nonequilibrium open quantum systems beyond weak coupling: a reaction coordinate approach}

\author{Khalak Mahadeviya}
\thanks{Corresponding author: \href{mailto:mahadevk@tcd.ie}{mahadevk@tcd.ie}}
\affiliation{School of Physics, Trinity College Dublin, College Green, Dublin 2, D02 K8N4, Ireland}
\author{Saulo V. Moreira}
\affiliation{School of Physics, Trinity College Dublin, College Green, Dublin 2, D02 K8N4, Ireland}
\author{Sheikh Parvez Mandal}
\affiliation{Departamento de Física - CIOyN, Universidad de Murcia, Murcia E-30071, Spain}
\author{Mahasweta Pandit}
\affiliation{Departamento de Física - CIOyN, Universidad de Murcia, Murcia E-30071, Spain}
\author{Javier Prior    }
\affiliation{Departamento de Física - CIOyN, Universidad de Murcia, Murcia E-30071, Spain}
\author{Mark T. Mitchison}
\thanks{Corresponding author: \href{mailto:mark.mitchison@kcl.ac.uk}{mark.mitchison@kcl.ac.uk}}
\affiliation{School of Physics, Trinity College Dublin, College Green, Dublin 2, D02 K8N4, Ireland}
\affiliation{Department of Physics, King’s College London, Strand, London, WC2R 2LS, United Kingdom}

\begin{abstract}
We investigate current fluctuations in open quantum systems beyond the weak-coupling and Markovian regimes, focusing on a coherently driven qubit strongly coupled to a structured bosonic environment. By combining full counting statistics with the reaction coordinate mapping, we develop a framework that enables the calculation of steady state current fluctuations and their temporal correlations in the strong-coupling regime. Our analysis reveals that, unlike in weak coupling, both the average current and its fluctuations exhibit nonmonotonic dependence on the system-environment interaction strength. Notably, we identify a regime where current noise is suppressed below the classical thermodynamic uncertainty bound, coinciding with enhanced anticorrelations in quantum jump trajectories and faster system relaxation. We further show that these features are linked to nonclassical properties of the reaction coordinate mode, such as non-Gaussianity and quantum coherence. Our results provide new insights and design principles for controlling current fluctuations in quantum devices operating beyond the standard weak-coupling paradigm.
\end{abstract}

\maketitle

\section{Introduction}

Open quantum systems far from equilibrium support currents of particles or energy, which can fluctuate significantly compared to their mean values~\cite{Esposito2009, Landi2024, Bruderer_2014}. These current fluctuations are important for several reasons: they control the tradeoff between power and efficiency in heat engines~\cite{Shiraishi2016, Pietzonka2016}, they limit the precision of parameter estimation in nonequilibrium settings~\cite{Tsang2013, Gammelmark2013, Kiilerich2014, Macieszczak2016, Albarelli2018, Yang2023, Cabot2024, prech2024optimaltimeestimationclock,Khandelwal2025,Mihailescu2025}, and they carry important information about system properties, e.g., particle statistics~\cite{Glidic2023,Ruelle2023,Iyer2024} or dissipative phase transitions~\cite{Kewming2022,Matsumoto2025,FlindtPhysRevLett.110.050601}. The universal features of current fluctuations have come under renewed scrutiny in recent years, with the discovery of general bounds such as the thermodynamic~\cite{barato_thermodynamic_2015, gingrich_dissipation_2016,Horowitz2020} and kinetic~\cite{Garrahan2017, terlizzi_kinetic_2018} uncertainty relations. Of particular interest is the potential of open quantum systems to violate these classical uncertainty relations, as demonstrated theoretically by numerous case studies~\cite{Krzysztof2018,brandner_thermodynamic_2018, Macieszczak2018, Agarwalla2018,Kalaee2021,Rignon-Bret2021,Prech2023} as well as via the derivation of looser, quantum bounds~\cite{Guarnieri2019,Hasegawa2020,hasegawa_thermodynamic_2021, Hasegawa2023a, VanVu2022, Timpanaro2023, Prech2024a, Macieszczak2024,Moreira2025,  Brandner2025,Blasi2025,Palmqvist2025}. This raises the tantalizing prospect of exploiting quantum resources to reduce current fluctuations, allowing for more reliable thermal machines~\cite{Krzysztof2018,Liu2019,Rignon-Bret2021, Timpanaro2025, Almanza-Marrero2025} or precise timekeeping~\cite{Meier2025}.

One way to realise nonclassical current fluctuations is by moving to a regime of strong coupling between the system and environment, where non-Markovian effects can arise. This regime is natural for many open quantum systems, because boundary effects often dominate due to the small sizes involved. However, the description of current fluctuations in strongly coupled, nonlinear open quantum systems typically requires sophisticated techniques such as perturbative nonequilibrium Green functions~\cite{Wang2014}, tensor-network simulations~\cite{Popovic2021, mandal2025}, or pseudomode approaches~\cite{Brenes2023,Bettmann2025}, whose predictive power comes at the cost of complexity which may obscure physical intuition. An appealing method in this regard is the reaction-coordinate (RC) mapping, in which the most significant environmental mode is incorporated as part of the system, while the remaining modes are traced out within a Born-Markov approximation~\cite{Iles-Smith2014,Iles-Smith2016,Nazir2018}. The RC mapping can be understood as the first step of a more general transformation of a noninteracting environment into a one-dimensional chain~\cite{Prior2010}, with the advantage that the RC captures the most important environmental properties explicitly and simply by a single mode. This approach has been widely used to model quantum thermodynamic processes at strong system-reservoir coupling, albeit mostly at the level of average heat and work exchanges~\cite{Strasberg2016a,Newman2017,Anto-Sztrikacs2021a, Anto-Sztrikacs2022a, Anto-Sztrikacs2023,Colla_2025,Brenes2025}. Notably, Shubrook et al.~\cite{Shubrook_2025} recently used the RC method to study the fluctuations of heat transferred during an equilibration process.


{In contrast to Ref.~\cite{Shubrook_2025} which examines transient equilibration process through Redfield dynamics,} we employ the RC mapping to investigate the fluctuations of currents sustaining a nonequilibrium steady state (NESS) beyond the weak-coupling regime. We consider a minimal model of a two-level system (qubit) driven by a coherent field and dissipating energy into a structured bosonic reservoir with a peaked spectral density (Sec.~\ref{sec:setup}), which can be mapped onto a dissipative Jaynes-Cummings model via the RC mapping (Sec.~\ref{sec:RC}). Here, we work within the rotating-wave approximation (RWA), which limits our study to moderate coupling strengths that are nonetheless well beyond the validity of the standard weak-coupling master equation. 
In this approximation, heat is transferred via the exchange of excitations (quanta) whose total number is conserved, and we therefore focus on the full counting statistics of this excitation current, as described in Sec.~\ref{sec:fcs}. 

Within our model, we find an interesting nonmonotonic dependence of the current noise on the system-environment coupling strength, induced by the structured environment. In particular, we show that the current noise dips below the classical TUR bound when the system-environment coupling is comparable to the Rabi frequency of the drive (Sec.~\ref{sec:current_fluctuations}). To understand this behavior, we exploit the fact that the RC mapping represents the original (possibly non-Markovian) setup via an extended system with Markovian dynamics, i.e.,~a Markovian embedding~\cite{woods_mappings_2014}. This allows us to apply the powerful conceptual framework of quantum-jump unraveling, whereby the excitation current can be modeled as a sequence of detector ``clicks'' whose statistical properties are inherited from the underlying quantum dynamics~\cite{GarrahanPhysRevLett.104.160601,Landi2024}. In particular, we find that the dip in the current noise is associated with strong anticorrelations between subsequent clicks (Sec.~\ref{sec:correlations}), a hallmark of the nonlinearity induced by qubit-RC hybridisation. We also show that the system's relaxation rates---encoded by the Liouvillian eigenvalues---are maximal at this point in parameter space. Finally, we characterise nonclassical features of the environment by quantifying the non-Gaussianity~\cite{Genoni2008}, second-order coherence~\cite{glauber1963quantum}, and coherence in the Fock basis~\cite{Baumgratz14} of the RC mode (Sec.~\ref{sec:nonclassicality}). We find that, although these quantifiers are correlated with reduced current fluctuations, there is no sharp correspondence between nonclassicality of the environment state and quantum TUR violations, in accordance with previous studies~\cite{Rignon-Bret2021, Liu2021, Prech2023}. 

Our results provide design principles for suppressing current fluctuations using strong coupling to a structured environment. Our work also demonstrates how the RC mapping facilitates a physically transparent analysis of current fluctuations beyond the weak-coupling regime, by exploiting the Markovian nature of the embedding. A similar approach could be applied to study more complex environmental structures or transient fluctuations~\cite{Blasi2024} using sophisticated embeddings developed in recent years~\cite{Mascherpa2020,Brenes2020, Elenewski2021, Lacerda2023, purkayastha2021periodically, purkayastha2022periodically, Menczel2024}.

\section{Model}
\label{sec:model}
\subsection{Coherently driven qubit}
\label{sec:setup}

We consider a coherently driven qubit coupled to a bosonic thermal bath at temperature $T$. The system Hamiltonian can be expressed as
\begin{equation}
    \label{eq:Driven_qubit_sys}
    \hat{H}_{\rm S}(t) = \omega_q\hat{\sigma}_+\hat{\sigma}_- + 2\Omega {\cos}(\omega_d t)\hat{\sigma}_x
\end{equation}
where $\omega_q$ is the qubit transition frequency, $\Omega$ is the Rabi frequency, $\omega_d$ is the drive frequency, and $\hat{\sigma}_{x, y, z, +, -}$ denote Pauli matrices. Throughout this work, we set $\hbar = 1$ and $k_{\rm B} =1$. Additionally, the bath Hamiltonian is given by
\begin{equation}
    \label{eq:original_bath}
    \hat{H}_{\rm B} = \sum_k {\nu}_k \hat{c}_k^\dagger \hat{c}_k,
\end{equation}
where $\hat{c}_k^\dagger (\hat{c}_k)$ are bosonic creation (annihilation) operators for modes with frequencies $\nu_k$, with $[\hat{c}_k,\hat{c}_{k^{\prime}}^\dagger]=\delta_{k,k^{\prime}}$.
By including the interaction between system and bath, the total Hamiltonian reads
\begin{equation}\label{eq:driven_qubit_total}
    \hat{H}(t) = \hat{H}_{\rm S}(t)  +\hat{H}_{\rm B} +\hat{H}_{\rm SB}.
\end{equation}
The system-bath interaction Hamiltonian is given by
\begin{equation}
    \label{eq:sb-int-ham}
    \hat{H}_{\rm SB} = \sum_k h_k (\hat{\sigma}_+ +\hat{\sigma}_-)( \hat{c}_k^\dagger +\hat{c}_k ),
\end{equation}
where $h_k$ is the coupling strength between the qubit and the bath frequency modes $\nu_k$.
In effect, the interaction between the system and the bath can be captured via the spectral density, $S(\omega)= {2\pi}\sum_k |h_k|^2 \delta(\omega-\nu_k)$~\cite{Leggett1987}.  
We assume the Drude-Lorentz form of spectral density
corresponding to a strong coupling between the system and bath frequency modes centered around $\omega_c$,
\begin{equation}
    \label{eq:spec-drude-lorentz}
    S(\omega) = \frac{4\omega \alpha \lambda^2 \omega_c^2 }{(\omega^2 - \omega_c^2)^2 + (2 \pi \alpha \omega_c \omega )^2}.
\end{equation}
The amplitude of the spectral function, $\lambda$, is related to the interaction strength, $h_k$, as $\lambda^2 = \sum_k |h_k|^2$, and its width $\omega_c \alpha$ is determined by a dimensionless parameter $\alpha$.

\begin{figure}[t]
    \centering
    \includegraphics[width=0.9\linewidth]{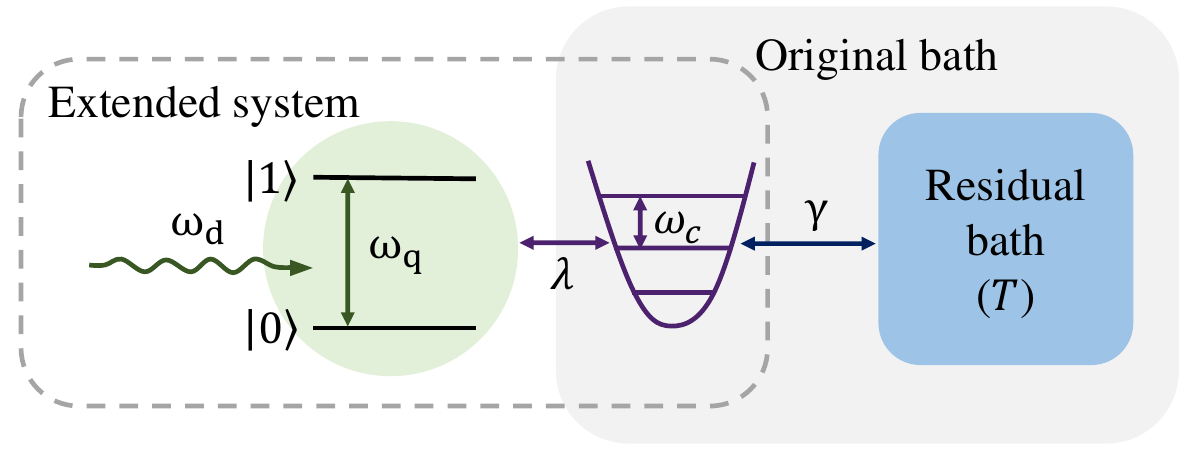}
    \caption{Reaction coordinate (RC) mapping for a coherently driven qubit strongly coupled to its environment. The RC mapping incorporates a harmonic oscillator mode into the extended system. $|0\rangle$ and $|1\rangle$ denote the qubit eigenstates in the $\sigma_z$ basis, with transition frequency $\omega_q$. The qubit couples to a drive at $\omega_d$ and the RC mode with frequency $\omega_{c}$ and coupling $\lambda$. The extended qubit-RC system interacts weakly ($\gamma$) with the residual bath at temperature $T$.}
    \label{fig:qubit-rc}
\end{figure}

\begin{figure}[t]
    \centering
    \includegraphics[width=0.8\linewidth]{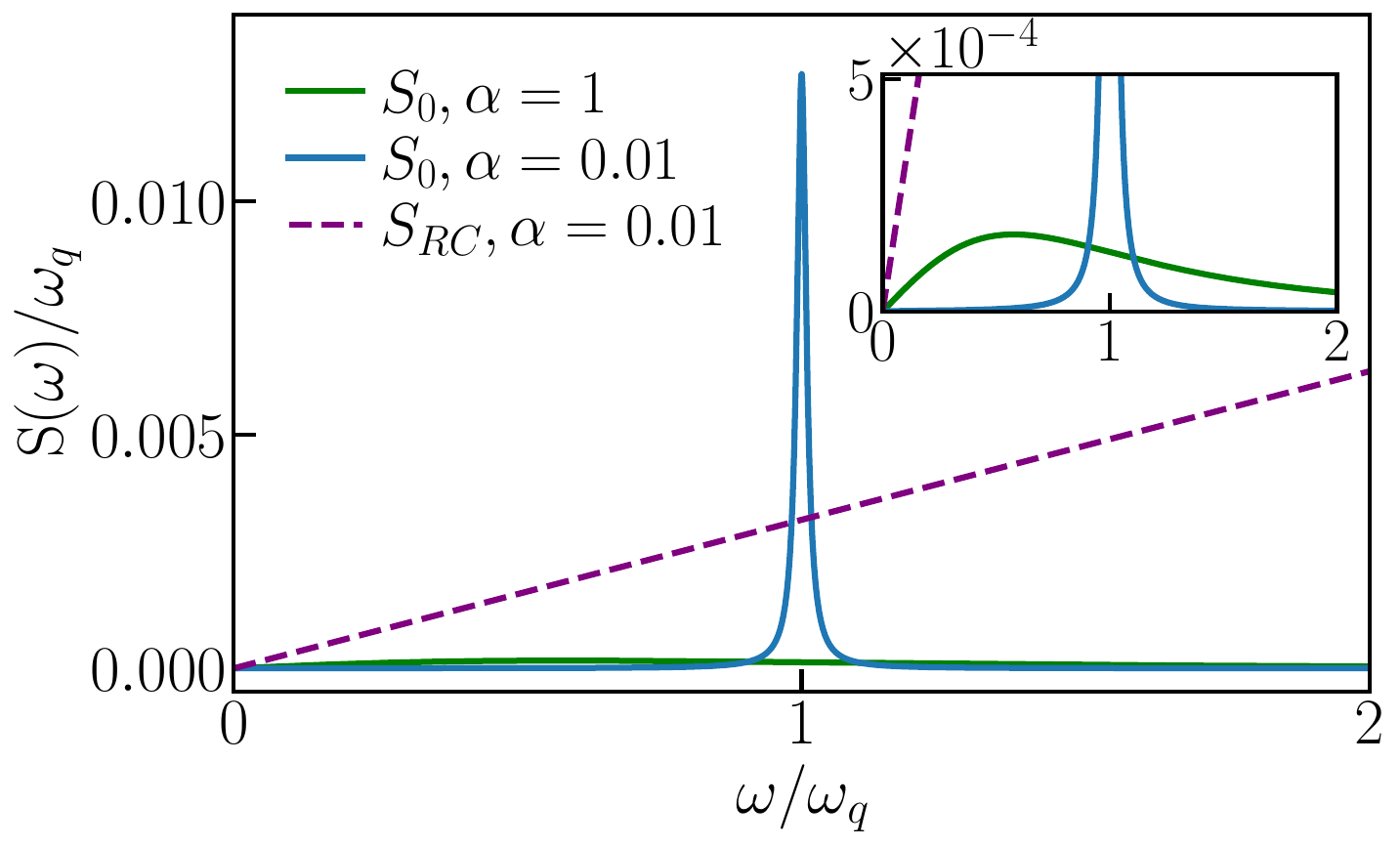}
    \caption{Drude-Lorentz spectral density with central frequency $\omega_c = \omega_q$ and $\lambda = 0.02 \omega_q$ for different widths $\alpha = 0.01$ (blue) and $1$ (green). The sharply peaked curve for $\alpha = 0.01$ represents the strong coupling to the bath and the dashed purple curve represents the Ohmic spectral density function for the residual environment after RC mapping.}
    \label{fig:sepctral-densities}
\end{figure}

In order for the Markov approximation to be valid, the system relaxation time ($\tau_{\rm S}$) has to be much larger than the bath correlation time ($\tau_{\rm B}$), i.e., $\tau_{\rm S} \gg \tau_{\rm B}$.  For the above spectral density, $\tau_{\rm B}$ is related to the inverse of the width of the spectral function $ \omega_c \alpha$. In this context, the Markovian regime is then defined by,
\begin{equation}
    \label{eq:markov-approx}
    \omega_c \alpha \gg \lambda.
\end{equation}
This condition results in flatter spectral densities as shown in the green curve in Fig.~\ref{fig:sepctral-densities} with $\lambda = 0.02\omega_q$, $\omega_c = \omega_q$ and $\alpha = 1$, and the system dynamics for this case can be described within the standard Lindblad master equation formalism.

\subsection{Reaction coordinate mapping}
\label{sec:RC}
In this work, we focus on investigating particle current fluctuations beyond the weak coupling regime. 
In the model described above, the strong system-bath coupling, $\lambda \gtrsim \omega_c \alpha$, corresponds to a narrow spectral density centered around the qubit transition frequency $\omega_q$, as shown in the blue curve in Fig.~\ref{fig:sepctral-densities}.
This means that we go beyond the Markovian regime, and the standard Lindblad formalism is not valid to describe the system dynamics. However, we can redefine the system by incorporating a collective mode of the environment, as illustrated in Fig.~\ref{fig:qubit-rc}. By constructing an \emph{extended system} that couples weakly to the residual environment, we are able to derive a Lindblad master equation governing the dynamics of the extended system. This can be achieved by implementing the reaction coordinate (RC) mapping method developed in Refs.~\cite{Iles-Smith2014,Iles-Smith2016, Nazir2018}.

The application of the RC mapping to the total Hamiltonian in Eq.~(\ref{eq:driven_qubit_total}) allows us to write the mapped Hamiltonian as
\begin{equation}
    \label{eq:RC_driven_qubit_total}
    \begin{aligned}
        \hat{H}^{\prime}(t) = \hat{H}_{\rm S}(t) + \hat{H}_{\rm RC} +\hat{H}_{\rm S,RC} +\hat{H}_{\mathrm{B}^{\prime}} + \hat{H}_{\rm RC,B^{\prime}}.
    \end{aligned}
\end{equation}
The RC and system-RC interaction Hamiltonian are defined as
\begin{equation}
    \label{eq:sys-RC-int}
    \hat{H}_{\rm RC} =  \omega_{c} \hat{a}^\dagger \hat{a}, \quad \hat{H}_{\rm S,RC} = \lambda \hat{\sigma}_x (\hat{a}^\dagger + \hat{a}),  
\end{equation}
with the bosonic creation (annihilation) operators $\hat{a}^\dagger(\hat{a})$ corresponding to the RC mode.
The residual bath is described by modes $\hat{b}_k^\dagger(\hat{b}_k)$ with frequency $\omega_k$ with Hamiltonian,
\begin{equation}
    \label{eq:residual-bath}
    \hat{H}_{\mathrm{B}^{\prime}} = \sum_k {\omega}_k \hat{b}_k^\dagger\hat{b}_k,
\end{equation}
and the interaction Hamiltonian between the residual bath and the RC mode is given by,
\begin{equation}
    \label{eq:res-bath-RC-int}
     \hat{H}_{\rm RC,B^{\prime}} = \sum_k f_k (\hat{a}^\dagger+ \hat{a})( \hat{b}_k^\dagger + \hat{b}_k).
\end{equation}
We identify the effective system Hamiltonian as $\hat{H}_{\rm ES} = \hat{H}_{\rm S}(t) + \hat{H}_{\rm RC} +\hat{H}_{\rm S,RC}$. 

The RC mode is defined such that $\lambda(\hat{a}^\dagger+\hat{a}) = \sum_{k}h_k (\hat{c}^\dagger_k+\hat{c}_k)$, and RC mode frequency $\omega_c^2 = \lambda^{-2} \sum_k \nu_k |h_k|^2 $. 
As discussed in Refs.~\cite{Iles-Smith2016,Iles-Smith2014},  under the RC mapping the original bath spectral density in Eq.~\eqref{eq:spec-drude-lorentz} transforms to an Ohmic form,
\begin{equation}
    \label{eq:spec-ohmic-rc}
    S_{\rm RC}(\omega) = {2\pi}\sum_k |f_k|^2 \delta(\omega - \omega_k) = \alpha \omega e^{-\omega/\Lambda},
\end{equation}
for the residual bath. We note that this correspondence between the spectral densities holds exactly only in the limit $\Lambda \to \infty$. In this work, however, we approximate it by setting a large but finite value of $\Lambda$. This is important because it ensures the convergence of all the terms in the derivation of the master equation discussed below. 

To derive the master equation for the extended system, we move to a rotating frame defined by the unitary operator ${\hat{U}(t) = \exp(i\omega_d\hat{N}_{\rm tot}t)}$, where $\hat{N}_{\rm tot}=\hat{\sigma}_+\hat{\sigma}_- + \sum_k \hat{c}_k^\dagger \hat{c}_k = \hat{\sigma}_+\hat{\sigma}_- + \hat{a}^\dagger \hat{a} + \sum_k \hat{b}_k^\dagger \hat{b}_k$ is the total excitation number operator. We note that our choice of rotating frame defined by $\hat{U}(t)$ is equivalent for both the original and RC-mapped description. In the rotating frame, the RC mapped Hamiltonian in Eq.~\eqref{eq:RC_driven_qubit_total} is transformed as $ {\tilde{H}^{\prime} = \hat{U}(t) \hat{H}^{\prime} \hat{U}^\dagger(t) + i\frac{d  \hat{U}(t)}{dt} \hat{U}^\dagger(t)}$. Under the rotating wave approximation (RWA), we can neglect fast oscillating terms to get the following Hamiltonian, with only remaining coupling terms that conserve the excitation number,
\begin{equation}
    \label{eq:RC_driven_qubit_rot}
    \begin{aligned}
        \tilde{H}^{\prime} =& ~\Delta_q\hat{\sigma}_+\hat{\sigma}_- + \Omega (\hat{\sigma}_+ +\hat{\sigma}_-) + \lambda (\hat{\sigma}_+\hat{a} + \hat{\sigma}_ -\hat{a}^\dagger ) \\
        & + \Delta_{c} \hat{a}^\dagger \hat{a}  +\sum_k f_k (\hat{a}^\dagger \hat{b}_k + \hat{a} \hat{b}_k^\dagger ) + \sum_k {\delta}_k \hat{b}_k^\dagger \hat{b}_k,
    \end{aligned}
\end{equation}
where $\Delta_q =  \omega_q - \omega_d$, $\Delta_c = \omega_c - \omega_d$, and $\delta_k = \omega_k - \omega_d$. 
In this frame, the residual bath spectral density is transformed into $\tilde{S}_{\rm RC}(\omega) =  S_{\rm RC}(\omega + \omega_d)$. 
The RWA holds when the drive frequency $\omega_d$ is near resonant to both the qubit transition frequency, $\omega_q$, and the RC frequency, $\omega_c$, i.e., $\Delta_q, \Delta_c \ll \omega_d$, and the Rabi coupling $\Omega$ and the qubit-RC coupling $\lambda$ 
are weak enough and satisfy $\Omega, \lambda \ll {\omega_d}$.

The total Hamiltonian in Eq.~\eqref{eq:RC_driven_qubit_rot} commutes with the total excitation number operator $\hat{N}_{\rm tot}$, which represents the total number of energy quanta in the system and bath. This conservation law is a consequence of the RWA, which dictates that energy exchange between subsystems occurs via the exchange of these excitations. We therefore focus on the excitation current between the system and the bath, rather than the energy current, in the following.

Now, we trace out the residual bath to derive a quantum master equation in Gorini-Kossakowski-Sudarshan-Lindblad (GKSL) form, starting from the total Hamiltonian in the rotating frame in Eq.~\eqref{eq:RC_driven_qubit_rot}. We work within a local approach to dissipation, which is justified so long the drive strength $\Omega$, coupling $\lambda$, and detunings $\Delta_{q,c}$ are small compared to the inverse bath correlation time $\tau_{\rm B^{\prime}} = \max\{\Lambda^{-1},T^{-1}\}$, which is dictated here by the inverse temperature of the bath due to the very large UV cutoff $\Lambda$. As discussed in Refs.~\cite{Rivas_2010, Hofer2017, Trushechkin2021, Potts_2021,Schnell2025}, we can then treat the bath spectral function near frequency ${\omega_d}$ as approximately constant on the scale of the small energy splittings induced by $\Omega, \lambda, \Delta_{q,c}$. 
We thus arrive at the following local Lindblad master equation {(see Appendix~\ref{ap:me_derivation_sketch} for the derivation)},
\begin{equation}
    \label{eq:RC-LME-driv-qbit}
     \frac{d{\hat{\rho}}}{dt}  = \mathcal{L}\hat{\rho} = -i[\tilde{H}_{\rm ES},\hat{\rho}] + \gamma(n_{\rm B}+1)\mathcal{D}[\hat{a}]\hat{\rho} + \gamma n_{\rm B} \mathcal{D}[\hat{a}^\dagger]\hat{\rho}.
\end{equation}
with $\gamma = {S}_{\rm RC} ({\omega_d})$, and the bosonic occupation number $n_{\rm B} \equiv n_{\rm B}({\omega_d}) = (\exp({\omega_d}/T)-1)^{-1}$ for the residual bath at temperature $T$. Here, $\tilde{H}_{\rm ES} = \Delta_q\hat{\sigma}_+\hat{\sigma}_- + \Omega (\hat{\sigma}_+ +\hat{\sigma}_-) + \Delta_{c} \hat{a}^\dagger \hat{a} +  \lambda (\hat{\sigma}_+\hat{a} + \hat{\sigma}_ -\hat{a}^\dagger )$ is the rotating frame Hamiltonian and $\hat{\rho}$ is the density matrix of the extended system, which includes the driven qubit and the RC mode. 

The above master equation is derived within the Born-Markov approximation, which imposes that the residual bath relaxes much faster than the extended system relaxation timescale ($\tau_{\rm ES}\sim \gamma^{-1}$). This approximation, like our use of a local GKSL equation, assumes an approximately flat spectral function ${S_{\rm RC}(\omega)n_{\rm B}(\omega)}$ around the relevant transition frequencies of the system. In our case, this is justified so long as $\Omega,\lambda,\Delta_{q,c} \ll \omega_d, T$ and $n_{\rm B}\lesssim 1$. Furthermore, validity of the Born approximation is limited to weak coupling between the extended system and environment, requiring the coupling constants $f_k$ to be weak. This assumption is also crucial for the validity of the RWA above. Although the Ohmic residual spectral density in Eq.~\eqref{eq:spec-ohmic-rc} implies that $f_k$ can become unboundedly large for large $\omega_k$, this assumption is justified since we coarse-grain over times much longer than the bath correlation time, so that these high-frequency contributions average out to zero.

From here on, we refer to Eq.~(\ref{eq:RC-LME-driv-qbit}) as reaction coordinate Lindblad master equation (RC-LME). In the long-time limit, the extended system dynamics governed by the RC-LME evolves towards a time-independent state, referred to as the nonequilibrium steady state (NESS). The NESS, denoted by $\rho_{\rm ss}$, satisfies
\begin{equation}
    \mathcal{L}\rho_{\rm ss} = 0,
\end{equation}
and represents the stationary solution of the RC-LME. {Note that we have numerically checked that the Liouvillian in Eq.~\eqref{eq:RC-LME-driv-qbit} has a single zero eigenvalue for all parameters considered in this work, which guarantees a unique steady state.}

\subsection{Full counting statistics}
\label{sec:fcs}

\subsubsection{Excitation current into the original bath}
\label{sec:fcs-original-bath}
Given the RC-LME in Eq.~(\ref{eq:RC-LME-driv-qbit}), we are interested in studying the steady state excitation current into a strongly coupled environment. To compute the first and second cumulants of this excitation current we employ the method of full counting statistics (FCS)~\cite{Landi2024}. 

Starting from the original description, note that under the rotating frame transformation introduced in the previous section, the Hamiltonian takes the form $\tilde{H} = \tilde{H}_{\rm S} + \tilde{H}_{\rm B} + \hat{H}_{\rm SB}$, where $\tilde{H}_S = \Delta_q \hat{\sigma}_+ \hat{\sigma}_- + {\Omega} (\hat{\sigma}_+ + \hat{\sigma}_-)$ and $\tilde{H}_{\rm B} = \sum_k (\nu_k - \omega_d)\hat{c}_k^\dagger \hat{c}_k$. Within this description, we define the excitation transfer using the two-point measurement scheme, formulated in the measurement basis of the original bath excitation number,  $\hat{N}_{\rm B} = \sum_k \hat{c}_k^\dagger \hat{c}_k$. Under this protocol, we consider projective measurements on the initial system-bath state, $\hat{\rho}_{\rm tot}(0) = \hat{\rho}{(0)} \otimes \hat{\rho}_{E}$, and on the unitarily evolved state, $\hat{\rho}_{\rm tot}(t) = \exp(-i\tilde{H}t)\hat{\rho}_{\rm tot}(0)\exp(i\tilde{H}t) $, at a later time $t$. The difference $N_{\rm B}$ between the outcomes of such measurements corresponds to the net number of excitations exchanged with the bath, during time $t$.
The probability distribution $P(N_{\rm B},t)$ of observing $N_{\rm B}$ at time $t$ can therefore be constructed by repeating the protocol many times.
Note that $P(N_{\rm B},t)=\mathrm{Tr} [\hat{\rho}(N_{\rm B},t)]$, where $\hat{\rho}(N_{\rm B},t)$ denotes the (unnormalised) conditional state associated to the ensemble of protocols in which exactly $N_{\rm B}$ net excitations are transferred after time $t$. Accordingly, the unconditional state reads $\hat{\rho}(t) = \sum_{N_{\rm B}=0}^\infty \hat{\rho}(N_{\rm B},t)$.

To compute the full statistics of the counting variable $N_{\rm B}$, it is convenient to introduce the characteristic function of $P(N_{\rm B},t)$, defined as its Fourier transform,
\begin{equation}
\label{moment_generating_function}
    M(\chi,t) = {\rm Tr }[\hat{\rho}(\chi,t)] = \sum_{N_{\rm B}} e^{i\chi N_{\rm B}} {\rm Tr } [\hat{\rho}(N_{\rm B},t)].
\end{equation}
Here, $\chi$ is called a \emph{counting field}. The counting-field dependent system state $\hat{\rho}(\chi,t)$ can be shown to obey a generalised quantum master equation of the form \cite{Esposito2009, Landi2024},
\begin{equation}
    \label{eq:gqme}
    \frac{d\hat{\rho}(\chi,t)}{dt} = \mathcal{L}_\chi \hat{\rho}(\chi,t), 
\end{equation}
with the counting-field dependent \emph{tilted Liouvillian} $\mathcal{L}_\chi$, and $\hat{\rho}(\chi,0)= \hat{\rho}(0)$. The solution of such a generalised QME is given by $\hat{\rho}(\chi,t) = e^{\mathcal{L}_\chi}\hat{\rho}(0)$.

Now, we can write the cumulant generating function $\phi(\chi,t)$ as,
\begin{align}
    \label{eq:cumulant_gen_1}
        \phi(\chi,t) &=\ln\big[M(\chi,t)\big] \notag\\
        & = \ln \Big[{\rm{Tr}}\big[e^{\mathcal{L}_\chi t}\hat{\rho}(0)\big]\Big].
\end{align}
With this, the $n^{th}$ cumulant of the net excitation at time $t$, $\llangle N_{\rm B}(t) ^{n} \rrangle$,  can be computed as,
\begin{equation}
    \label{eq:cumulants_1}
    \llangle N_{\rm B}(t) ^{n} \rrangle = (-i \partial_\chi)^n \phi(\chi,t) \bigg|_{\chi = 0}.
\end{equation}
As for the steady state statistics, in the long time limit, this cumulant generating function takes the following asymptotic form~\cite{Landi2024},
\begin{equation}
    \label{eq:cumulant_gen_2}
    \phi(\chi,t) \simeq \theta_0(\chi)t,
\end{equation}
where $\theta_0(\chi)$ is the leading eigenvalue of the tilted Liouvillian $\mathcal{L}_\chi$, i.e., the eigenvalue with the largest real part. Thus, using Eq.~\eqref{eq:cumulants_1} and Eq.~\eqref{eq:cumulant_gen_2}, we are able to compute cumulants of net excitations $N_{\rm B}(t)$ in the long time limit. 
By defining the scaled cumulant generating function,
\begin{equation}
    \phi(\chi) = \underset{t\to\infty}{\rm{lim}} \partial_t\phi(\chi,t) =\theta_0(\chi),
\end{equation}
the scaled cumulants of the excitations in the long time limit are given by,
\begin{equation}
    \label{eq:scaled_cumulants}
     \partial_t\llangle N_{\rm B}(t) ^{n} \rrangle = (-i \partial_\chi)^n\theta_0(\chi)\bigg|_{\chi = 0}.
\end{equation}

In order to employ the FCS framework to calculate the statistics of the steady state excitation current in our setup, we are required to derive a counting-field dressed generalised master equation (GME) of the form in Eq.~\eqref{eq:gqme}. To this end, we start from the rotating frame Hamiltonian introduced earlier, $\tilde{H} = \tilde{H}_{\rm S} + \tilde{H}_{\rm B} + \hat{H}_{\rm SB}$, with $\tilde{H}_S = \Delta_q \hat{\sigma}_+ \hat{\sigma}_- + {\Omega} (\hat{\sigma}_+ + \hat{\sigma}_-)$ and $\tilde{H}_{\rm B} = \sum_k (\nu_k - \omega_d)\hat{c}_k^\dagger \hat{c}_k$. To account for excitation transfer into the original bath, we introduce a counting field $\chi$ by transforming the total Hamiltonian as $\tilde{H} \to e^{i\chi \hat{N}_{\rm B}/2}\tilde{H}e^{-i\chi \hat{N}_{\rm B}/2}$, to get the \emph{tilted Hamiltonian}, 
\begin{equation}
    \label{eq:tlt-ham-og}
    \hat{H}_\chi = \tilde{H}_{\rm S} + \tilde{H}_{\rm B} + e^{i\chi \hat{N}_{\rm B}/2}\hat{H}_{\rm SB}e^{-i\chi \hat{N}_{\rm B}/2}.\
\end{equation}
The counting field $\chi$ in the interaction term keeps track of the excitation exchange at the original system-bath boundary. 

Within the RC mapping, the interaction term transforms as, $\hat{H}_{\rm SB} \to \hat{H}_{\rm S,RC}$. Hence, the \emph{RC mapped tilted Hamiltonian} is given by~\cite{Brenes2023},
\begin{align}
    \label{eq:tlt-ham-rc}
    \hat{H}^{\prime}_{{\rm tot},\chi} = \tilde{H}_S + \tilde{H}_{\rm RC}  &+ e^{\frac{i}{2}\chi \hat{N}_{\rm B}}\hat{H}_{\rm S,RC}e^{-\frac{i}{2}\chi \hat{N}_{\rm B}} \notag\\ 
    &+ \tilde{H}_{\rm B^{\prime}}+\hat{H}_{\rm RC,B^{\prime}},
\end{align}
with $\tilde{H}_{\rm RC} = \Delta_c \hat{a}^\dagger \hat{a}$ and $\tilde{H}_{\rm B^{\prime}} = \sum_k \delta_k\hat{b}_k^\dagger \hat{b}_k$. Note that the original bath number operator $\hat{N}_{\rm B}$ also satisfies  $\hat{N}_{\rm B} = \sum_k \hat{c}_k^\dagger \hat{c}_k = \hat{a}^\dagger \hat{a} +\sum_k \hat{b}_k^\dagger \hat{b}_k$.
{Now, using the Baker-Campbell-Hausdorff (BCH) formula, the term $e^{i\chi \hat{N}_{\rm B}}\hat{H}_{\rm S,RC}e^{-i\chi \hat{N}_{\rm B}}$ simplifies to
\begin{equation}
\label{tilted_interaction_ham}
    \hat{H}_{\chi,{\rm S,RC}} = \lambda(e^{i\chi/2}~\hat{a}^\dagger\hat{\sigma}_- +e^{-i\chi/2} ~\hat{\sigma}_+\hat{a}).
\end{equation}}
By utilising the RC-mapped tilted Hamiltonian in Eq.~\eqref{eq:tlt-ham-rc}, we derive the \emph{RC-mapped tilted Liouvillian} under the same approximations as in the previous section, to get a generalised master equation (GME) of the form~\eqref{eq:gqme} with
\begin{align}
    \label{eq:tilt-H-liouv}
      \mathcal{L}_\chi \hat{\rho} 
        & = -i[\tilde{H}_{\rm ES, \chi},\hat{\rho}]_\chi +\gamma(n_{\rm B}+1) \mathcal{D}[\hat{a}] \hat{\rho} + \gamma n_{\rm B} \mathcal{D}[\hat{a}^\dagger]\hat{\rho},
\end{align}
with $\tilde{H}_{\rm ES, \chi} = \tilde{H}_S + \tilde{H}_{\rm RC} + \hat{H}_{\chi, S, RC}$ and $[\tilde{H}_{\rm ES, \chi},\hat{\rho}]_\chi = \tilde{H}_{\rm ES, \chi}\hat{\rho} - \hat{\rho}\tilde{H}_{\rm ES,-\chi}$. The RC mapped tilted Liouvillian $\mathcal{L}_\chi$, in the above GME with counting field $\chi$, accounts for the excitation exchange between the system and the RC mode. The full statistics for the corresponding excitation current can be calculated using Eq.~\eqref{eq:scaled_cumulants}.

\subsubsection{Excitation current into the residual bath}
\label{sec:fcs-residual-bath}

Similarly, we can derive another GME with a counting field $\chi^{\prime}$ that accounts for the excitation exchange between the extended system and the residual bath. This can be achieved by following the same two-point measurement protocol as described above, but now in the measurement basis of the residual bath excitation number $\hat{N}_{\rm B^{\prime}} = \sum_k \hat{b}_k^\dagger \hat{b}_k.$ The associated counting variable, which corresponds to the difference between the initial and final measurement outcomes, is denoted by $N_{\rm B^{\prime}}$. Within this protocol, we obtain the following RC-mapped tilted Hamiltonian,
\begin{align}
    \label{eq:tlt-ham-rc-res-bath}
    \hat{H}^{\prime}_{{\rm tot},\chi^{\prime}} = & ~\tilde{H}_S + \tilde{H}_{\rm RC}  +\hat{H}_{\rm S,RC} + \tilde{H}_{\rm B^{\prime}} \notag\\ 
    &+  e^{\frac{i}{2}\chi^{\prime} \hat{N}_{\rm B^{\prime}}}\hat{H}_{\rm RC,B^{\prime}}e^{-\frac{i}{2}\chi^{\prime} \hat{N}_{\rm B^{\prime}}}.
\end{align}
With this tilted Hamiltonian, we can derive the corresponding GME of the form,
\begin{equation}
    \label{eq:tlt-D-liouv}
    \mathcal{L}^{\prime}_{\chi^{\prime}} \hat{\rho} = -i[\tilde{H}_{\rm ES}, \hat{\rho}] 
    + \gamma (n_{\rm B} +1 )\mathcal{D}_{\chi^{\prime}}[\hat{a}] \hat{\rho} + \gamma n_{\rm B}\mathcal{D}_{-\chi^{\prime}}[{\hat{a}^\dagger}] \hat{\rho}, 
\end{equation}
with
\begin{equation}
    \label{eq:tlt-dissipator}
    \mathcal{D}_{\chi}[\hat{L}] \hat{\rho} = e^{i \chi} \hat{L} \hat{\rho}\hat{L}^\dagger - \frac{1}{2}\{\hat{L}^\dagger\hat{L},\hat{\rho}\}.
\end{equation}
The full statistics of the net excitation current into the residual bath, related to the counting variable $N_{\rm B^{\prime}}$, can be obtained from the cumulant generating function associated with the above GME.

\subsubsection{Equivalence of the two excitation currents}
\label{sec:fcs-equivalence}

Using the framework described above, we are now in a position to show that, in the long time limit, \emph{the statistics of the net excitations $N_{\rm B}$ exchanged between the system and original bath is identical to the statistics of net excitations $N_{\rm B^{\prime}}$ exchanged between the RC and the residual environment}. {Intuitively, this can be understood since most excitations entering the RC from the qubit will eventually dissipate into the residual bath or return to the qubit. Excitations temporarily stored in the RC will lead to corrections to the statistics of $N_{\rm B}(t)$ that do not scale with time and are thus irrelevant for the scaled cumulants~\eqref{eq:scaled_cumulants} in the limit $t\to\infty$.}

To formalise this equivalence, in Appendix~\ref{ap:FCS_equivalence}, using the following unitary superoperators,  
\begin{align}
    \label{eq:unitary-sup-2}
    & \mathcal{U}\hat{\rho} = e^{-\frac{i}{2} \chi \hat{a}^\dagger \hat{a}} \hat{\rho} e^{-\frac{i}{2} \chi \hat{a}^\dagger \hat{a}}, \notag \\ 
    & \mathcal{U}^\dagger\hat{\rho} = e^{\frac{i}{2} \chi \hat{a}^\dagger \hat{a}} \hat{\rho} e^{\frac{i}{2} \chi \hat{a}^\dagger \hat{a}},
\end{align}
we demonstrate that the tilted Liouvillian $\mathcal{L}_\chi$ (Eq.~\eqref{eq:tilt-H-liouv}) and $\mathcal{L}^{\prime}_{\chi^{\prime}}$ (Eq.~\eqref{eq:tlt-D-liouv}) are related as,
\begin{equation}
    \label{eq:FCS-equivalence}
    \mathcal{U}^\dagger\mathcal{L}_\chi \mathcal{U} = \mathcal{L}^{\prime}_{\chi}.
\end{equation}%

From Eq.~\eqref{eq:cumulant_gen_2}, we recall that in the long-time limit, all the cumulants of the net excitation are determined by the eigenvalues of the corresponding tilted Liouvillian. Since the leading eigenvalue $\theta_0(\chi)$ of the tilted Liouvillian is invariant under unitary transformation, both GMEs in Eq.~\eqref{eq:tlt-D-liouv} and Eq.~\eqref{eq:tilt-H-liouv} yield an identical cumulant generating function $\phi(\chi,t) = \theta_0(\chi)t$ in the long-time limit. This implies that the cumulants of the excitation current between the system and environment in the original description can be calculated directly from the GME in Eq.~\eqref{eq:tlt-D-liouv}.
From here on,
we drop the subscripts from $N_{\rm B}$ and $N_{\rm B^{\prime}}$ {wherever it will not cause confusion}, and refer to the net number of excitations as $N$.

{
\subsubsection{Thermodynamic quantities}
\label{sec:thermodynamics}

Before we can discuss the excitation current statistics in relation to the TUR, we also need to consider the rate of entropy production, which is associated with the flow of heat into the bath~\cite{Landi2021}. Here, we treat the irreversible exchange of energy with the residual bath as heat, $\dot{Q}$, while the energy absorbed from the coherent drive is treated as work or power, $\dot{W}$. Heat and work must sum to the internal energy change by the first law of thermodynamics. However, since we are dealing with a coupled, driven system, it is not \textit{a priori} obvious how to assign the internal energy, $U$, in a thermodynamically consistent way.

To solve this problem, we adopt the philosophy of Ref.~\cite{Potts_2021} and define the internal energy $U = \langle \hat{H}_{\rm TD}\rangle$ with the ``thermodynamic Hamiltonian''
\begin{equation}
    \label{thermodynamic_Hamiltonian}
    \hat{H}_{\rm TD} = \omega_d \left( \hat{\sigma}_+\hat{\sigma}_- + \hat{a}^\dagger \hat{a} \right).
\end{equation}
Writing the master equation~\eqref{eq:RC-LME-driv-qbit} as $d\hat{\rho}/dt = \mathcal{L}\hat{\rho} = -i [\hat{H}_{\rm ES},\hat{\rho}] + \mathcal{D}\hat{\rho}$, where the superoperator $\mathcal{D}$ encapsulates all dissipative processes, we then define
\begin{align}
\label{heat}
    & \dot{Q} = {\rm tr}(\hat{H}_{\rm TD} \mathcal{D}\hat{\rho}), \\
    \label{work}
    & \dot{W} = -i \,{\rm tr}(\hat{H}_{\rm TD} [\hat{H}_{\rm ES},\hat{\rho}]).
\end{align}
Eq.~\eqref{heat} assigns an amount of heat $\omega_d$ to each excitation absorbed from the residual bath, i.e.~$\dot{Q} = -\omega_d\partial_t\langle N_{\rm B^{\prime}}\rangle$, as shown in Appendix~\ref{ap:thermodynamic_hamiltonian}. This assumption is fully consistent with our local GKSL model derived in Appendix~\ref{ap:me_derivation_sketch}, which assumes that the residual bath cannot resolve energy differences on the order of the small splittings $\lambda, \Omega, \Delta_{q,c} \ll \omega_d, T$, since these are comparable in magnitude to the linewidth $ \gamma$. Meanwhile, Eq.~\eqref{work} is equivalent to the standard expression for the power in the lab frame $\dot{W} = \langle \partial_t \hat{H}(t)\rangle$~\cite{alickiQuantumOpenSystem1979} under the RWA, as shown in Appendix~\ref{ap:thermodynamic_hamiltonian}. 

Here, we are primarily concerned with the entropy production rate $\dot{\Sigma} = -\dot{Q}/T$, and not with the work or heat themselves. Nevertheless, our definitions are thermodynamically consistent in the following sense. In the absence of a drive ($\Omega = 0$), the system equilibrates to a thermal state with internal energy given by Eq.~\eqref{thermodynamic_Hamiltonian}, i.e.~the steady state is $e^{-\beta \hat{H}_{\rm TD}}/Z$, consistent with the zeroth law. The first law is satisfied identically since $\dot{Q} + \dot{W} =  \dot{U} \equiv {\rm tr}(\hat{H}_{\rm TD} \mathcal{L}\rho)$. The second law,  $\dot{\Sigma} = -\dot{Q}/T \geq 0$, is satisfied numerically for all parameters considered in this work, providing a self-consistency check on our results. We note that alternative definitions of heat and work within reaction-coordinate mappings have recently been explored in Refs.~\cite{Colla_2025, Shubrook_2025}. These authors also conclude that entropy production within the RC approach should be associated with heat exchanged irreversibly with the residual bath.

}

\subsubsection{Current fluctuations and correlations}
\label{sec:current-statistics}

Finally, we can now define the average excitation current and diffusion coefficient in terms of the first and second scaled cumulants of the net excitation transfer $N(t)$ in the long-time limit,
\begin{equation}\label{Dt}
    J = \lim_{t\to\infty}\frac{d}{dt}{\rm E} [N(t)],  \quad D = \lim_{t\to\infty}\frac{d}{dt} \text{Var}[N(t)],
\end{equation}
where ${\rm E}[\bullet]$ denotes the expectation value (first cumulant) and ${\rm Var}[\bullet]$ denotes the variance (second cumulant). In terms of these quantities {and the entropy production rate $\dot{\Sigma} = -\dot{Q}/T$}, the thermodynamic uncertainty relation (TUR) reads
\begin{equation}
    \label{eq:TUR}
    \mathcal{Q} = \frac{D }{J^2}~\dot{\Sigma} \geq 2.
\end{equation}
This trade-off between steady state current fluctuations and entropy production holds for classical systems undergoing Markovian dynamics with local detailed balance~\cite{barato_thermodynamic_2015,gingrich_dissipation_2016,Horowitz2020}. {Combining Eqs.~\eqref{heat} and~\eqref{Dt}, we find $\dot{\Sigma} = - \omega_d J/T$ in steady state (see Appendix~\ref{ap:thermodynamic_hamiltonian}). The thermodynamic uncertainty ratio $\mathcal{Q}$ can thus be simplified to
\begin{equation}
     \label{eq:TUR-2}
     \mathcal{Q} = \frac{D}{J}~{\ln(n_{\rm B}^{-1}+1)}.
 \end{equation}
}
To obtain our results in Sec.~\ref{sec:results}, the average current $J$ and noise $D$ are computed from the leading eigenvalue of the tilted generator defined in Eq.~\eqref{eq:tlt-D-liouv}. However, to obtain further insight into the statistics of the excitation current, we also exploit the quantum-jump unraveling of full counting statistics. Here, the counting variable $N$ is interpreted in terms of the ``clicks'' of an ideal detector that efficiently monitors the emitted and absorbed quanta~\cite{Landi2024}. Since the dynamics of the extended qubit-RC system is Markovian, such a detector can be introduced without affecting the dynamics or current statistics. Therefore, even if such an ideal detector cannot be implemented in practice, it is a useful fiction that provides a time-resolved picture of current fluctuations, even in this non-Markovian setting. 

We define $N_+(t)$ (respectively, $N_-(t)$) as the total number of excitations that the RC emits into (absorbs from) the residual bath. The total excitation transfer up to time $t$ is therefore given by $N(t) = N_+(t) - N_-(t)$. The associated stochastic current is defined as $I(t) = dN/dt$, such that its expectation value ${\rm E}[I(t)] = J$ reproduces the average current defined in Eq.~\eqref{Dt}. Following Ref.~\cite{Landi2024}, the statistics of this current can be determined from the current superoperator
\begin{equation}
    \mathcal{J}\hat{\rho} = \sum_{k=\pm} \nu_k \hat{L}_k \hat{\rho} \hat{L}_k^\dagger.
    \label{eq:current_operator}
\end{equation}
Each term in the sum describes a possible quantum jump in the trajectory unravelling of the RC-LME~\eqref{eq:RC-LME-driv-qbit}. When a jump is detected in channel $k$, the counting variable $N$ is incremented by a weight $\nu_k$ and the conditional state is updated by the action of the jump operator $\hat{L}_{k}$. In this case, $\nu_+ = +1$ and $\hat{L}_+ = \sqrt{\gamma (1+n_{\rm B})} \hat{a}$ describe emission processes that increase $N$ by one unit, while $\nu_- = -1$ and $\hat{L}_- = \sqrt{\gamma n_{\rm B}}\hat{a}^\dagger$ describe absorption events that correspondingly decrease it. 

The average steady state current can be computed as
\begin{equation}
    \label{eq:avg-current-t}
    J = {\rm{Tr}}[\mathcal{J}\hat{\rho}_{\rm ss}].
\end{equation}
The stationary two-point correlation function between $I(t)$ and $I(t+\tau)$ is given by
\begin{align}
    F(\tau) &= \text{E}[\delta I(t) \delta I(t+\tau)]\notag \\     & =  K \delta(\tau) + \text{Tr}[\mathcal{J}e^{\mathcal{L}|\tau|}\mathcal{J}\hat{\rho}_{\rm ss}]-J ^2,     \label{eq:two-point-correlation-F}
\end{align}
where $\delta I(t) = I(t) - J$ denotes current fluctuations and $K = \sum_k \nu_k^2 {\rm Tr}[\hat{L}_k^\dagger \hat{L}_k\hat{\rho}_{\text{ss}}]$. In our case, since $\nu_k = \pm 1$, $K$ corresponds to the dynamical activity~\cite{Maes2017, Maes2020}, which quantifies the frequency of jumps. As shown in Ref.~\cite{Landi2024}, the noise can be written in terms of the two-point correlation function as
\begin{equation}
\label{noise_F}
    D = 2\int_0^\infty d\tau F(\tau).
\end{equation}
Features of the noise can therefore be understood in terms of correlations within the stochastic current, as we discuss in more detail below.

\section{Results} 
\label{sec:results}
\subsection{Current fluctuations beyond weak coupling}
\label{sec:current_fluctuations}

\begin{figure}
    \centering
    \includegraphics[width=\linewidth]{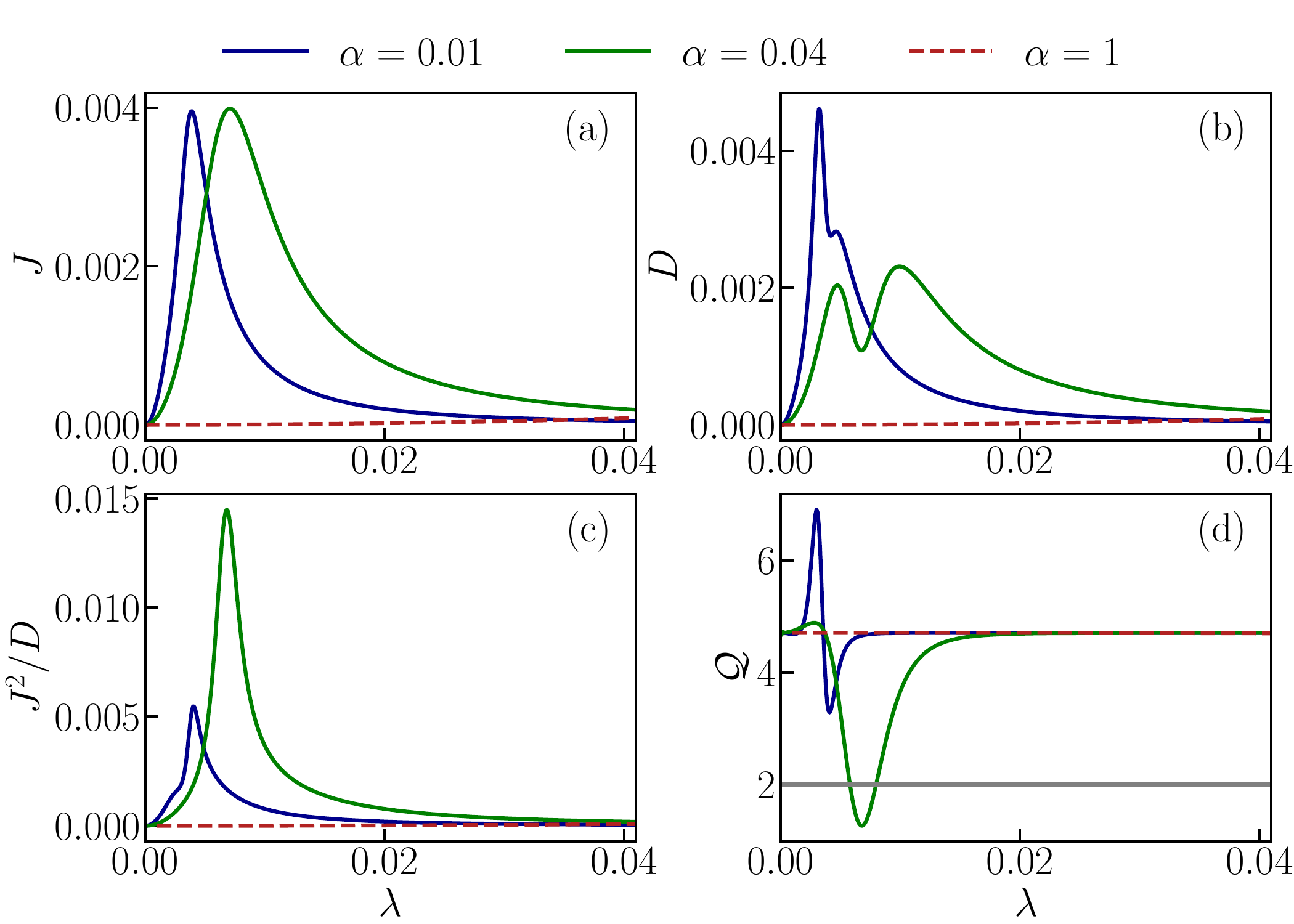}
    \caption{(a) Average excitation current $J$ into the bath, (b) fluctuations of the excitation current $D$, (c) signal-to-noise ratio (SNR) $J^2/D$, and (d) TUR ratio $\mathcal{Q}$ as functions of interaction strength $\lambda$ for different spectral density width $\alpha = 0.01,~ 0.04,$ and $1$. We set $\Delta_q=0$, $\Delta_c=0$, $\Omega = 0.005$, and $n_{\rm B} = 0.01$.}
    \label{fig:plot-cumulants-vs-int-wk-str}
\end{figure}

We now apply the methods developed in the previous section to compute the steady state excitation current and its cumulants for a coherently driven qubit strongly coupled to a bosonic environment, the dynamics of which is described by the RC-LME in Eq.~\eqref{eq:RC-LME-driv-qbit}. In Fig.~\ref{fig:plot-cumulants-vs-int-wk-str}, we present the average current defined in Eq.~\eqref{eq:avg-current-t}, noise (second cumulant), signal-to-noise ratio (SNR), and thermodynamic uncertainty ratio (TUR) as functions of the interaction strength $\lambda$. The results are shown for different spectral widths $\alpha$ in the Drude-Lorentz spectral density. In all our calculations, we take $\omega_c=1$ as the unit of energy, and we consider two values of $\alpha = 0.01$ and $0.04$, both in the strong-coupling regime. The weak-coupling case, $\alpha = 1$, which yields a relatively flat spectral density, is also considered as a benchmark. For the weak-coupling case, the heat current cumulants are computed using the standard Lindblad master equation and full counting statistics~\cite{Landi2024}. {All numerical calculations are performed in a truncated bosonic Hilbert space with dimension $d_{\rm RC} = 30$ for the RC mode; we have checked that all results are converged with respect to increases in $d_{\rm RC}$.}

We observe a clear qualitative difference between the strong and weak coupling regimes. For $\alpha = 1$, both the average current $J$ and noise $D$ increase monotonically with $\lambda$. In contrast, for smaller $\alpha$ values (blue and green curves), the average current exhibits a nonmonotonic dependence on $\lambda$, reaching a peak at intermediate coupling strengths and decreasing toward zero for large $\lambda$. The noise follows a similar trend to the current for $\alpha = 0.01$. However, for $\alpha = 0.04$, the noise displays a pronounced dip around the same $\lambda$ values where the current peaks. Notably, this dip coincides with the parameter regime where the classical TUR is violated, which occurs only in the strongly coupled cases. Despite similar current magnitudes near the peak, the SNR is notably higher for $\alpha = 0.04$, indicating that the TUR violation reflects a genuine reduction in fluctuations.

\begin{figure}
    \centering
    \includegraphics[width=\linewidth]{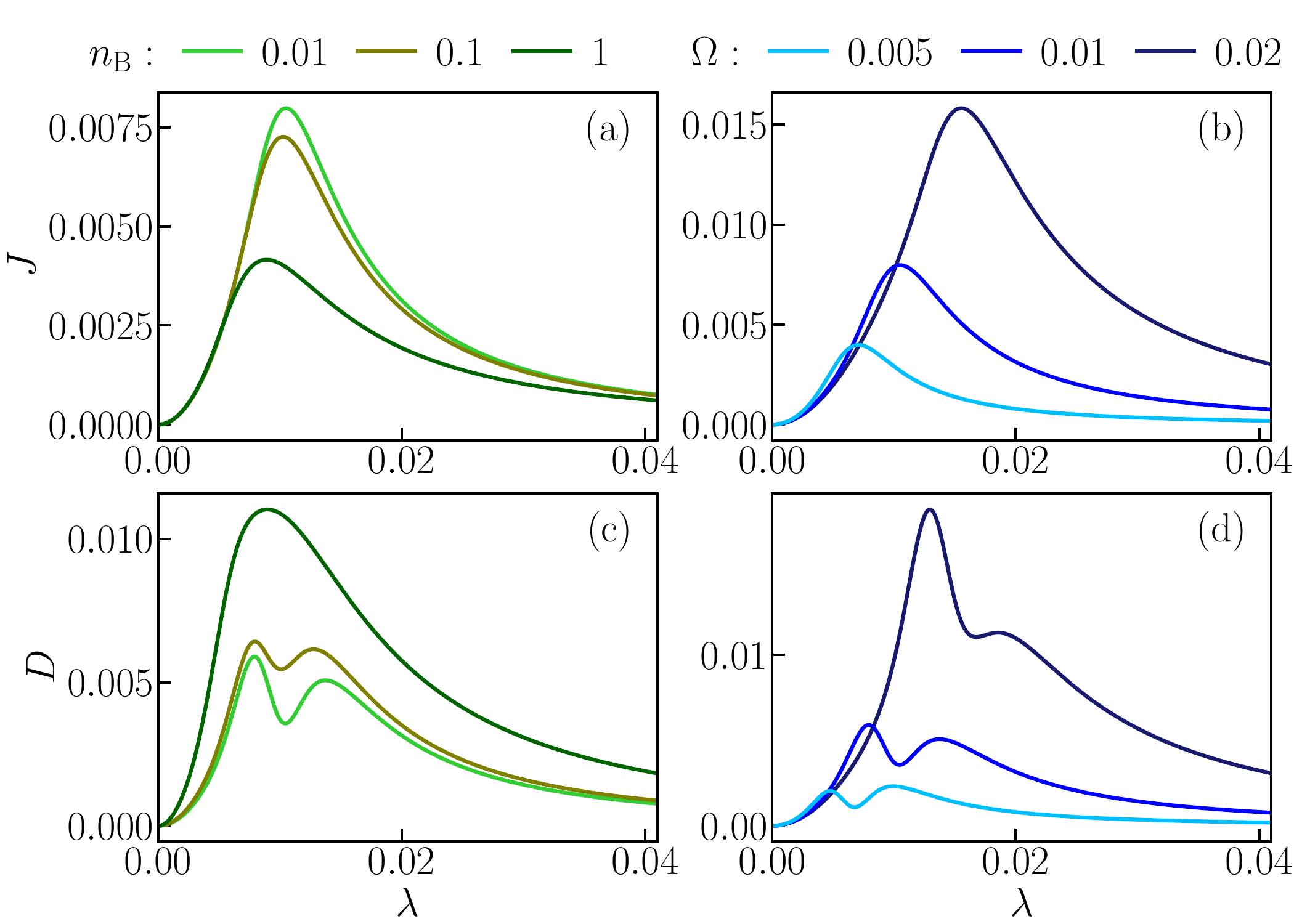}
    \caption{ Average excitation current $J$ into the bath and fluctuations of the excitation current $D$ as functions of interaction strength $\lambda$, in (a) and (c) for different temperatures leading to bath occupation $n_{\rm B} =  0.01,~ 0.1$, and $1$, with fixed $\Omega =0.01$, and in (b) and (d) for different drive strengths $\Omega = 0.005,~ 0.01$, and $0.02$, with fixed $n_{\rm B} =0.01$. We set $\Delta_q=0$ and $\Delta_c=0$.}
    \label{fig:plot-cuur-noise-vs-int-rc}
\end{figure}

We now focus on the case $\alpha = 0.04$, and plot the average heat current into the environment and the noise in Fig.~\ref{fig:plot-cuur-noise-vs-int-rc}, for different fixed values of the drive strength $\Omega$ and bath occupation $n_{\rm B}$. We first examine the dependence on bath temperature for a fixed drive strength $\Omega = 0.005$, shown in panels~\ref{fig:plot-cuur-noise-vs-int-rc}(a) and~\ref{fig:plot-cuur-noise-vs-int-rc}(c). As $n_{\rm B}$ decreases, we see an increase in the mean heat current and a reduction in noise, with the maximum current and minimum noise occurring at $n_{\rm B} = 0.01$, corresponding to the lowest bath temperature. Notably, the nonmonotonic features in the noise, characterised by distinct peaks and a trough, are most pronounced at this temperature, and gradually disappear as the temperature increases, vanishing for $n_{\rm B} \sim 1$. The enhanced current at low temperatures suggests that the system relaxes to the lower excited state more efficiently in this regime.

Next, we fix the bath occupation at $n_{\rm B} = 0.01$ and investigate the behavior of the current cumulants as a function of $\lambda$, for different values of the drive strength $\Omega$, as shown in Figs.~\ref{fig:plot-cuur-noise-vs-int-rc}(b) and~\ref{fig:plot-cuur-noise-vs-int-rc}(d). We consider three values: $\Omega = 0.005$, $0.01$, and $0.02$. The mean current increases with increasing drive strength, which is consistent with an overall increase in jump activity at stronger driving. However, this also leads to an increase in current fluctuations, with the peaks and troughs in the noise becoming less pronounced at higher $\Omega$. These features are most clearly resolved at the weakest drive.

\subsection{Current correlations}
\label{sec:correlations}

In order to deepen our understanding of the features shown in Fig.~\ref{fig:plot-cuur-noise-vs-int-rc}, we exploit the relation between the noise and the current auto-correlation function in Eq.~\eqref{noise_F}. We focus on the regular part of the correlation function by subtracting the singular term, $K\delta(\tau)$,  from Eq.~\eqref{eq:two-point-correlation-F},
\begin{equation}
    \label{eq:current-correlation-C}
    \begin{split}
        C(\tau) &\equiv F(\tau) - K \delta(\tau) \\
        &= \text{Tr}[\mathcal{J}e^{\mathcal{L}|\tau|}\mathcal{J}\hat{\rho}_{\rm ss}] - J^2.
    \end{split}
\end{equation}
Since the term $K\delta(\tau)$ is related to temporally uncorrelated fluctuations (shot noise)~\cite{Landi2024}, Eq.~\eqref{eq:current-correlation-C} quantifies contributions from temporal correlations between different quantum jumps. For example, when the jumps are totally uncorrelated, $F(\tau) = K\delta(\tau)$, whereas  $C(\tau)=0$.
Furthermore, we note that $C(\tau)$ can be related to the long-time diffusion coefficient as
\begin{equation}
    \label{eq:D-K-Corr}
    D-K =  \frac{1}{2}\int_{0} ^{\infty} C(\tau) \, d \tau.
\end{equation}
Once again, since the term $K$ is due to white noise, $D-K$ genuinely captures the contribution from temporal correlations. This quantity was previously introduced and studied in Ref.~\cite{Matsumoto2025} in connection with dissipative phase transitions. 

\begin{figure}
    \centering
    \includegraphics[width=\linewidth]{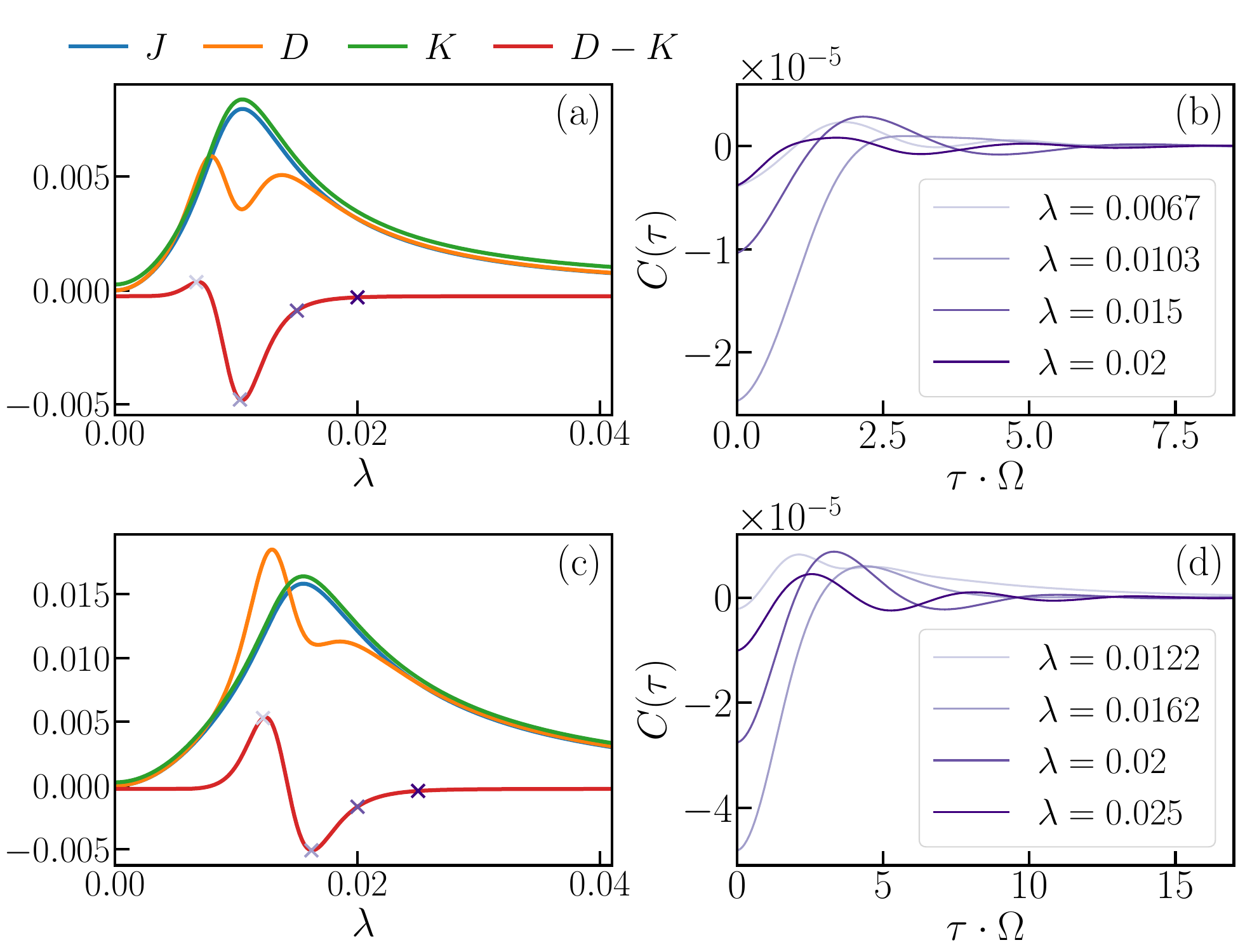}
    \caption{Average excitation current $J$, diffusion coefficient $D$, dynamical activity $K$, and $D-K$ as functions of interaction strength $\lambda$, in (a) and (c) for different drive strengths $\Omega = 0.01$ and $\Omega = 0.02$. In (b) and (d), current correlation functions $C(\tau)$ for selected $\lambda$ values marked by crosses in $D-K$ plots in panels (a) and (c), respectively.}
    \label{fig:plot-D-K-curr-corr}
\end{figure}
In Fig.~\ref{fig:plot-D-K-curr-corr}, we plot $D-K$ and $D$ as functions of the interaction strength $\lambda$, for $\Omega = 0.01$ in Fig.~\ref{fig:plot-D-K-curr-corr}(a), and $\Omega = 0.02$ in Fig.~\ref{fig:plot-D-K-curr-corr}(c). 
We see that $D-K$ follows the same trend as $D$, with peaks and troughs coinciding for the same values of $\lambda$. Such nonmonotonic behavior of $D$ and $D-K$ results from temporal correlations, as evidenced by Eq.~\eqref{eq:D-K-Corr}, with peaks (troughs) associated with correlated (anticorrelated) jumps.
It is therefore instructive to look at the behavior of the correlation function in Eq.~\eqref{eq:current-correlation-C} for values of $\lambda$ corresponding to peaks, troughs, alongside some other points, indicated in Figs.~\ref{fig:plot-D-K-curr-corr}(a)-(c).
We plot these correlation functions in
 Fig.~\ref{fig:plot-D-K-curr-corr}(b), for $\Omega = 0.01$, and Fig.~\ref{fig:plot-D-K-curr-corr}(d), for $\Omega = 0.02$.

We note that the correlation functions $C(\tau)$ shown in Fig.~\ref{fig:plot-D-K-curr-corr}(b)-(d) display an oscillatory behavior, starting at negative values for $C(0)$ and eventually converging to zero for large $\tau$. We highlight that at the troughs, in both cases, $C(0)$ reaches its most negative value, and the period of the oscillations of $C(\tau)$ is the largest, corroborating the predominance of anticorrelated jumps and, overall, antibunched statistics. This means that, near the troughs of $D$, the probability of observing a second jump is smallest at zero time delay after the first one ($\tau=0$). Such antibunching in a bosonic mode is a hallmark of nonlinear dynamics, indicating that strong hybridisation with the qubit prevents multiple excitations from being created in the RC (analogous to the photon blockade effect~\cite{Birnbaum2005a}).

\subsection{Relaxation timescales}

\begin{figure}
    \centering    \includegraphics[width=\linewidth]{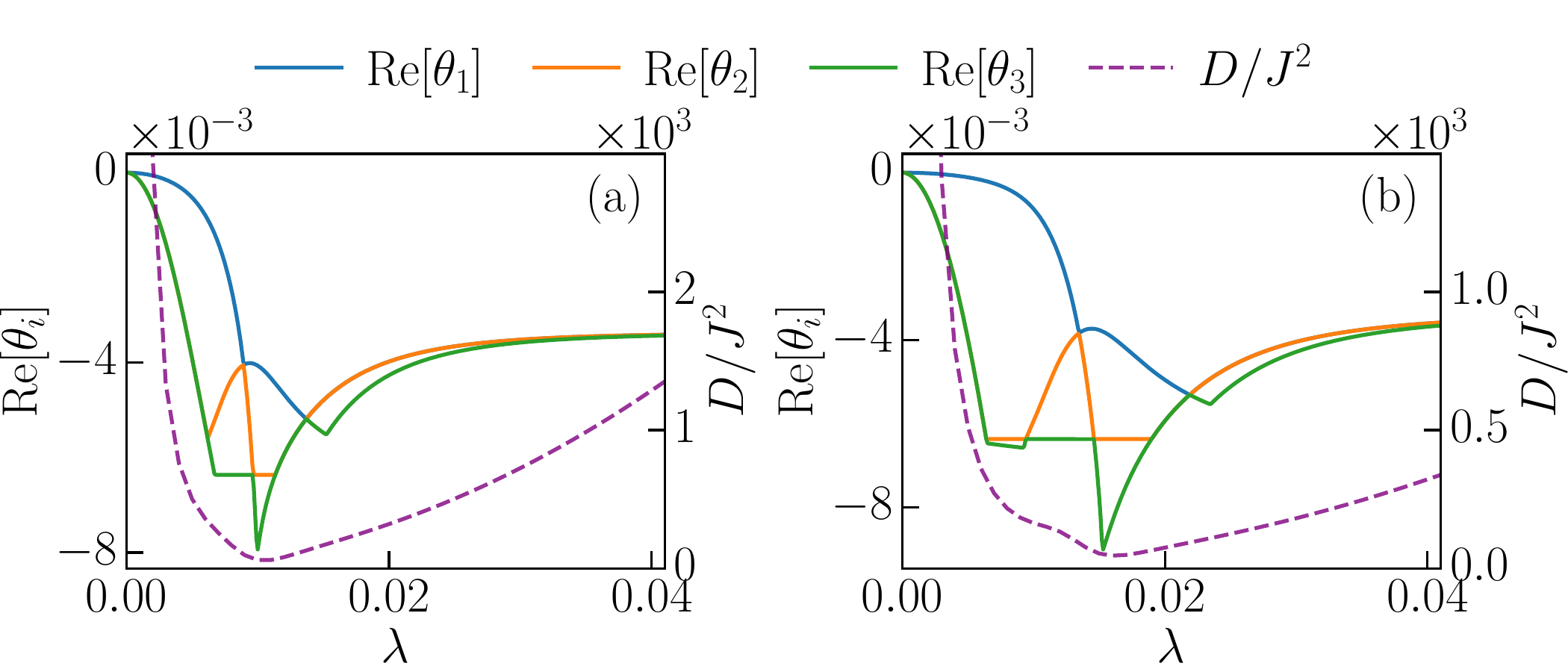}
    \caption{Real part of the first three nonzero Liouvillian eigenvalues $\theta_1$, $\theta_2$, $\theta_3$, and $D/J^2$ as functions of interaction strength $\lambda$, in (a) and (b) for different drive strengths $\Omega = 0.01$ and $\Omega = 0.02$. }
    \label{fig:plot-liouv-spec}
\end{figure}
As discussed in several previous works~\cite{Kewming2022, Landi2024, Matsumoto2025}, current fluctuations are associated with an intrinsic timescale of the system. In particular, consider the time-averaged stochastic current over an observation window $t$,
\begin{equation}
    \label{t_average_I}
    \bar{I}(t) = \frac{1}{t}\int_0^t dt^{\prime}\, I(t^{\prime}),
\end{equation}
which is an unbiased estimator of $J$ since ${\rm E}[\bar{I}(t)]=J$. The timescale $D/J^2$ is precisely the time $t$ taken for the signal-to-noise ratio of the estimator $\bar{I}(t)$ to reach one~\cite{Landi2024}. Thus, we expect the points in parameter space with minimal current noise to be associated with fast regression of current fluctuations back to the mean. 

To quantify these timescales more precisely, we consider the spectrum of the Liouvillian $\mathcal{L}$ of the extended system, Eq.~\eqref{eq:RC-LME-driv-qbit}. We note that the stationary state of the system satisfies $\mathcal{L}\rho_{\text{ss}} = 0$, and is therefore associated to the zeroth eigenvalue $\theta_0 = 0$. Furthermore, it is known that the real parts of the eigenvalues $\theta_{i>0}$, which satisfy ${\rm Re}[\theta_i] < 0$, determine the rates with which the system exponentially relaxes towards the steady state and with which correlation functions decay~\cite{Landi2024}. Note that this simple relation between relaxation timescales and the Liouvillian spectrum of the extended system is possible here due to the Markovian nature of the embedding.  

Here, we focus on the first three nonzero eigenvalues, which are plotted in Fig.~\ref{fig:plot-liouv-spec}(a), for $\Omega = 0.01$, and Fig.~\ref{fig:plot-liouv-spec}(b), for $\Omega=0.02$ [the same as in Figs.~\ref{fig:plot-D-K-curr-corr}(a)-(c)]. We see that the real parts of the eigenvalues become increasingly negative in the region where current fluctuations are minimal, with the lowest three eigenvalues reaching their minimum in the region where $D/J^2$ is lowest (dashed lines in Fig.~\ref{fig:plot-liouv-spec}). Thus, the reduction in current fluctuations due to strong coupling is correlated with faster relaxation of the \textit{extended} qubit-RC system.

\subsection{TUR violation and nonclassicality in RC mode}\label{sec:nonclassicality}

An important feature of the RC mapping description is that the state of a part of the environment is accessible.  
This is a notable advantage of our approach, as it enables us to probe different aspects of nonclassicality in a part of the environment itself, namely the RC, and connect it quantitatively to violation of the classical TUR. Although TUR violations have been extensively studied in relation to nonclassical features of the system state, such as quantum coherence and entanglement \cite{Kalaee2021, Prech2023}, the nonclassicality of the environmental degrees of freedom has not been considered yet in this regard. 

\begin{figure}[t]
    \centering
    \includegraphics[width=\linewidth]{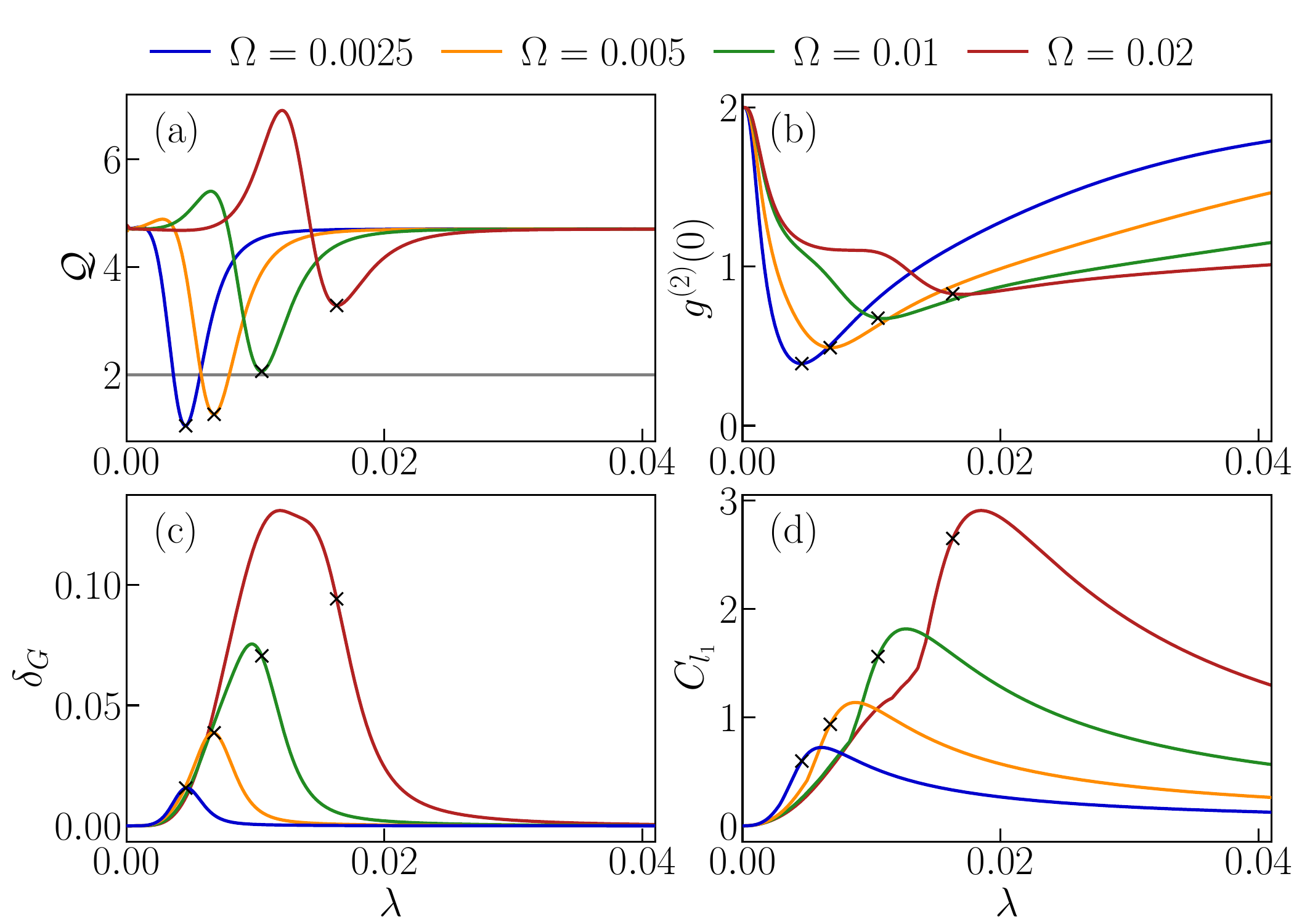}
    \caption{Comparison of the behavior of the TUR ratio $\mathcal{Q}$ (a) with second-order coherence $g^{(2)}(0)$ (b), non-Gaussianity $\delta_{G}$ (c), and $l_1$-coherence $\mathcal{C}_{l_1}$ of the reaction coordinate state (d) versus $\lambda$ for fixed $\alpha=0.04$, $n_{\rm B} = 0.01$, and $\Omega= \{0.0025,~ 0.005,~ 0.01,~ 0.02 \}$. The crosses represent the minimum TUR values.}
    \label{RC_TUR}
\end{figure}

In Fig.~\ref{RC_TUR}(a), we plot the TUR ratio $\mathcal{Q}$ as a function of $\lambda$, for different values of $\Omega$ and observe violations in two of the cases. We indicate, with black crosses, the points where each curve attains its minimum TUR value. First, we consider the zero-delay second-order coherence \cite{glauber1963quantum}  of the RC mode given by
\begin{equation}
    g^{(2)}(0)= \frac{\langle \hat{a}^{\dagger} \hat{a}^{\dagger} \hat{a} \hat{a} \rangle}{\langle \hat{a}^{\dagger} \hat{a}\rangle^2}.
\end{equation}
We plot $ g^{(2)}(0)$  in Fig.~\ref{RC_TUR}(b). Values $g^{(2)}(0) < 1$ at any given time indicate \emph{anticorrelation} of RC excitations. We see that the dips in $g^{(2)}(0)$ consistently coincide with the minima of $\mathcal{Q}$. Despite the fact that TUR violation corresponds to the largest dips in $g^{(2)}(0)$, anticorrelated behavior occurs even when the TUR bound is not violated, reflecting the fact that $\mathcal{Q}$ also depends on entropy production and model-dependent prefactors. Interestingly, we note that violations of the classical TUR bound occur very close to the interaction strength $\lambda$ where strongest anticorrelated nature ($g^{(2)}(0) < 1$) of the mode coincides with the largest values of its non-Gaussianity $\delta_G = S_{\text{vN}}(\tau) - S_{\text{vN}}(\rho)$, defined as the relative entropy distance between the RC state $\rho=\text{Tr}_{\rm S}[\rho_{\rm ss}]$ and its closest Gaussian state $\tau$ \cite{Genoni2008}, with $ S_{\text{vN}}(x) = -\text{Tr} [x \ln x ]$ being the von Neumann entropy.
Here, we use the fact that
\begin{equation}
    S_{\text{vN}}(\tau) =  \frac{\nu+1}{2}\ln \Big(\frac{\nu+1}{2}\Big) - \frac{\nu-1}{2}\ln \Big(\frac{\nu-1}{2}\Big)
\end{equation}
where $\nu$ is the symplectic eigenvalue of the covariance matrix of $\rho$ \cite{ weedbrook2012gaussian, serafini2023quantum}. We plot non-Gaussianity $\delta_G$ versus $\lambda$ in Fig.~\ref{RC_TUR}(c). Since $\delta_G \geq 0$, with equality if and only if the state is Gaussian, large $\delta_G$ implies divergence from any Gaussian description. Lastly, in Fig.~\ref{RC_TUR}(d), we consider a coherence quantifier, namely the $l_1$-coherence $\mathcal{C}_{l_1}=\sum_{i\neq j}|\rho_{ij}|$ of the RC mode in the Fock basis ~\cite{Baumgratz14}, and plot it as a function of $\lambda$. We observe coherence consistently peaks close to the values of $\lambda$ corresponding to the minima of $\mathcal{Q}$. 

It is important to point out that in the above analysis, we have considered two complementary measures of nonclassicality: the $l_1$-coherence in the Fock basis and non-Gaussianity. It is evident, for example, that any Fock state (which has $\mathcal{C}_{l_1}=0$) is a highly non-Gaussian state ($\delta_G > 0$), whereas a coherent state ($\delta_G = 0$) has large $\mathcal{C}_{l_1}$ in the Fock basis. Interestingly, the minima of $\mathcal{Q}$ occur precisely where these distinct signatures of nonclassicality intersect: where anticorrelation is the strongest, i.e., $g^{(2)}(0)$ is minimum, $\delta_G$ is maximal, and $\mathcal{C}_{l_1}$ is close to its peak. This observation highlights that several independent quantum traits in the environment, including both statistical (anticorrelation) and structural (non-Gaussianity and Fock-basis superposition) features of the RC mode, contribute to violations of the classical TUR.

\section{Discussion}

 In this work, we investigated the first and the second cumulants of excitation currents flowing through an open quantum system beyond the weak-coupling regime. In this direction, we begin by implementing the RC mapping framework and redefining the boundary between the system and environment by incorporating a collective environmental mode within an extended system. The interaction between the extended system and the residual bath can be described within the weak coupling approximation, allowing us to derive an RC-LME that governs the extended system dynamics. We show that the current statistics between different partitions, namely the original system-bath and the extended system-residual bath, are identical in the long-time limit. This enabled us to examine the features of current fluctuations and correlations in the strong coupling regime, utilizing the toolkit of Markovian dynamics.

 We find significantly different behaviors in the strong and weak-coupling regimes. While the average current, $J$, and noise, $D$, increase monotonically with the interaction strength in the weak-coupling regime, both quantities behave nonmonotonically in the strong-coupling regime. In addition, for the strongly-coupled cases, a dip in $D$ occurs in the same regime where a higher SNR is observed and the violation of the TUR occurs, thereby showing a suppression of current fluctuations. We showed that such nonmonotonic behavior of the noise results from strong temporal anticorrelations in the current. Furthermore, we have seen that larger negative values of the real part of eigenvalues of the Liouvillian, which imply faster relaxation of the system dynamics, are associated with a decrease in $D$. This is corroborated by the fact that the noise-to-signal ratio, $D/J^2$, which captures the timescale of the decay of fluctuations, also displays smaller values where a reduction in current fluctuations is observed. Finally, we investigated TUR violations in relation to complementary measures of nonclassicality in the environmental degrees of freedom. We observe that the suppression of current fluctuations is associated with the simultaneous occurrence of two different nonclassical features of the environment: namely, non-Gaussianity and Fock-basis coherence in the RC mode.

 {Our results are immediately relevant for experimental settings in which the system-bath coupling is mediated by a high-quality resonator. Such a configuration is possible, for example, in circuit-QED platforms, where a superconducting qubit strongly couples to a microwave resonator mode that is damped by transmission lines acting as a broadband environment~\cite{BlaisRevModPhys.93.025005}. However, the RC mapping is not limited to qubits, and may also be fruitfully applied in other settings where the environment has sharp, resonant spectral features, e.g.~the vibrational environment of photosynthetic energy-harvesting complexes~\cite{Adolphs2006,Chin2013,limVibronicOriginLonglived2015}.} 
  
It is important to note that our results are obtained within the RWA and the local GKSL master equation, where energy transfer between the system and the environment proceeds via the exchange of well-defined energy quanta. This essentially means that the heat cumulants are proportional to the cumulants of the net excitation transfer $N$, which we have shown are the same for the original and residual baths in the long-time limits. 
This greatly simplifies the thermodynamic analysis and singles out the current between the residual bath and the extended system as the natural choice to describe heat flow~\cite{Colla_2025, Shubrook_2025}. {This simplification comes at the cost of reducing the range of system-bath couplings that we can consider, as compared to the analysis of Shubrook et al.~\cite{Shubrook_2025}, which is based on a Redfield equation and thus avoids any kind of secular approximation. Nevertheless, we have shown that we can use the RC mapping to analyze qualitatively different features of current fluctuations beyond the weak-coupling regime, while still respecting the condition $\lambda \ll \omega_q$ needed for validity of the RWA.}

Within its regime of validity, we expect the RWA to give an adequate description up to small, fast-oscillating contributions that are neglected. However, for stronger system-bath couplings such that $\lambda \sim \omega_q$, or other more complex environmental spectral densities, the RWA will break down completely. In this regime, more sophisticated embeddings are required~\cite{Mascherpa2020, Brenes2020, Elenewski2021, Lacerda2023, purkayastha2021periodically, purkayastha2022periodically, Menczel2024} and additional theoretical complications will emerge, e.g., the nonequivalence of entropy flux between the original and residual baths~\cite{Lacerda2024, Shubrook_2025, Colla_2025}, and contributions to the heat current from interactions within the extended system~\cite{Lacerda2023, Bettmann2025}. Notwithstanding these subtleties, we hope that our results highlight the conceptual power of the Markovian embedding for understanding current fluctuations beyond weak system-environment coupling, and inspire further research along these lines deep into the strong-coupling regime.

\begin{acknowledgments}
We warmly acknowledge Kesha for the use of her computational resources.  M.T.M. is supported by a Royal Society University Research Fellowship. J.P. acknowledges support from grant QuantERA II program (Mf-QDS) that has received funding from the European Union’s Horizon 2020 research and innovation program under Grant Agreement No. 101017733 and from the Agencia Estatal de Investigación, with project code PCI2022-132915 funded by MICIU/AEI/10.13039/501100011033 and by the European Union NextGenerationEU/PRTR, grant TED2021-130578BI00, grant PID2021-124965NB-C21 (Quantera II cofund 2023 (Aqusend) Grant Agreement No. 101017733) and PCI2024-153474 funded by MICIU/AEI/10.13039/501100011033.".  This project is co-funded by the European Union (Quantum Flagship project ASPECTS, Grant Agreement No. 101080167) and UK Research and Innovation (UKRI). Views and opinions expressed are, however those of the authors only and do not necessarily reflect those of the European Union, Research Executive Agency or UKRI. Neither the European Union nor UKRI can be held responsible for them.
\end{acknowledgments}

\appendix

\setcounter{equation}{0}
\setcounter{figure}{0}
\setcounter{table}{0}
\makeatletter
\renewcommand{\theequation}{A\arabic{equation}}
\renewcommand{\thefigure}{A\arabic{figure}}

{\section{Master equation derivation}
\label{ap:me_derivation_sketch}

\subsection{Derivation of Eq.~\eqref{eq:RC-LME-driv-qbit}}\label{me_derivation}

In this appendix, for completeness, we outline the derivation of the local GKSL master equations used in the main text. We start from Eq.~\eqref{eq:RC_driven_qubit_rot}: the rotating-frame Hamiltonian after the reaction-coordinate mapping, including the system, RC mode, and residual bath. We move to an interaction picture with respect to $\tilde{H}_0 = \tilde{H}_{\rm ES} + \sum_k \delta_k \hat{b}^\dagger_k \hat{b}_k$, where  $\tilde{H}_{\rm ES} =\Delta_q\hat{\sigma}_+\hat{\sigma}_- + \Omega \hat{\sigma}_x + \Delta_{c} \hat{a}^\dagger \hat{a} +  \lambda (\hat{\sigma}_+\hat{a} + \hat{\sigma}_ -\hat{a}^\dagger )$, by the transformation 
\begin{equation}
\label{int_picture}
    \tilde{\rho}_{\rm tot}(t) = e^{i\tilde{H}_0t}\hat{\rho}_{\rm tot}(t)e^{-i\tilde{H}_0t},
\end{equation}
where $\hat{\rho}_{\rm tot}(t)$ refers to the total state of the system and bath in the original (rotating) frame and $\tilde{\rho}_{\rm tot}(t)$ its interaction-picture equivalent. The initial condition is taken to be the product state $\hat{\rho}_{\rm tot}(0) = \hat{\rho}(0) \hat{\rho}_{\rm B^{\prime}}$, where $\hat{\rho}_{\rm B^{\prime}}=e^{-\beta\hat{H}_{\rm B^{\prime}}}/Z_{\rm B^{\prime}}$ and $\hat{\rho}(t)$ is the state of the extended system. That is, we assume the residual bath is initially in thermal equilibrium (with respect to its Hamiltonian $\hat{H}_{\rm B^{\prime}}$ in Eq.
~\eqref{eq:residual-bath}) and uncorrelated with the extended system, which is justified by the weak residual coupling after the RC mapping~\cite{Iles-Smith2014,Nazir2018}. 

Following the standard derivation (e.g., see Refs.~\cite{breuerTheoryOpenQuantum2007,Rivas_2010}), we work at second order in the system-bath interaction $\hat{V} = \sum_k f_k (\hat{a}^\dagger \hat{b}_k + \hat{a} \hat{b}_k^\dagger )$ and invoke the Born-Markov approximation to find
\begin{equation}
\label{BM_master_equation}
    \frac{d\tilde{\rho}}{dt} = -\int_0^\infty d\tau\, {\rm tr}_{\rm B^{\prime}}[\tilde{V}(t), [\tilde{V}(t-\tau), \tilde{\rho}(t) \hat{\rho}_{\rm B^{\prime}} ]],
\end{equation}
where $\tilde{V}(t) = e^{i\tilde{H}_0t}\hat{V}e^{-i\tilde{H}_0t} = \sum_k f_k\tilde{a}(t)\tilde{b}^\dagger_k(t) + \rm h.c.$ and $\tilde{\rho} = {\rm tr}_{\rm B^{\prime}}[\tilde{\rho}_{\rm tot}]$ are respectively the Hamiltonian and the extended system state in the interaction picture. Eq.~\eqref{BM_master_equation} can be written explicitly as~\cite{breuerTheoryOpenQuantum2007,Rivas_2010}
\begin{align}
\label{me_eigenops}
    \frac{d\tilde{\rho}}{dt} & = \sum_{\omega,\omega^{\prime}} \left(\Gamma_-(\omega) e^{i(\omega^{\prime}-\omega)t}\left[\hat{a}(\omega)\tilde{\rho}\hat{a}^\dagger(\omega^{\prime}) -\hat{a}^\dagger(\omega^{\prime}) \hat{a}(\omega)\tilde{\rho}\right] \right. \notag \\ & + \left.  
    \Gamma_+(\omega) e^{i(\omega^{\prime}-\omega)t}\left[\hat{a}^\dagger(\omega)\tilde{\rho}\hat{a}(\omega^{\prime}) -\hat{a}^\dagger(\omega^{\prime}) \hat{a}(\omega)\tilde{\rho}\right]\right) +{\rm h.c.}
\end{align}
Above, the residual bath correlation functions are 
\begin{align}
\label{Gamma_-}
    \Gamma_-(\omega) & = \int_0^\infty d\tau\, e^{i\omega \tau} \sum_k |f_k|^2 \langle \tilde{b}_k(\tau) \tilde{b}^\dagger_k(0) \rangle_{\rm B^{\prime}} \notag \\ 
    & = \frac{1}{2} S_{\rm RC}(\omega_d + \omega) [1 + n_{\rm B}(\omega_d+\omega)],
\end{align}
\begin{align}
\label{Gamma_+}
    \Gamma_+(\omega) & = \int_0^\infty d\tau \, e^{i\omega \tau} \sum_k |f_k|^2 \langle \tilde{b}^\dagger_k(\tau) \tilde{b}_k(0) \rangle_{\rm B^{\prime}} \notag \\ 
    & = \frac{1}{2} S_{\rm RC}(\omega_d-\omega) n_{\rm B}(\omega_d-\omega),
\end{align}
where $\langle \bullet\rangle_{\rm B^{\prime}} = {\rm tr}[\bullet\hat{\rho}_{\rm B^{\prime}}]$ and we neglect the small Lamb-shift corrections associated with the imaginary components of $\Gamma_{\pm}(\omega)$. The ladder operators obey the relation
\begin{equation}
    \label{lowering_op}
    [\tilde{H}_{\rm ES}, \hat{a}(\omega)] = -\omega \hat{a}(\omega),
\end{equation}
where $\omega$ are transition frequencies (eigenvalue differences) of $\tilde{H}_{\rm ES}$.

To obtain a local GKSL equation, we assume that all of these transition frequencies are small compared to the drive frequency $\omega_d$ and temperature $T$, and approximate the correlation functions as constants in the vicinity of $\omega = 0$  (corresponding to $\omega = \omega_d$ in the lab frame). That is, we Taylor expand to zeroth order:
\begin{equation}
    \label{local_approx}
    \Gamma_{\pm}(\omega) = \Gamma_{\pm}(0) + \omega \Gamma_{\pm}^{\prime}(0) + \ldots  \approx \Gamma_\pm(0).
\end{equation}
Plugging this into Eq.~\eqref{me_eigenops}, returning to the original frame by inverting the transformation~\eqref{int_picture}, and using the identity $\sum_\omega \hat{a}(\omega) = \hat{a}$, we arrive at Eq.~\eqref{eq:RC-LME-driv-qbit}. This is a good approximation so long as higher-order terms in the Taylor expansion of $\Gamma_\pm(\omega)$ are negligible. In particular, comparing the first- and zeroth-order terms,
\begin{equation}
    \frac{\omega \Gamma_{\pm}^{\prime}(0)}{ \Gamma_\pm(0)} = \mathcal{O}(\omega/\omega_d) + \mathcal{O}(\omega/T) + \mathcal{O}(\omega n_{\rm B}(\omega_d)/T),
\end{equation}
we conclude that higher-order corrections can be consistently neglected so long as $\omega \ll \omega_d$, $\omega \ll T$ and $n_{\rm B}(\omega_d) \lesssim 1$. The transition frequencies scale as the detunings and couplings $\omega \sim \Delta_{q,c}, \lambda, \Omega \ll \omega_d, T$, where we always take $T\lesssim \omega_d$, so the approximation is justified for all parameters considered in this work. Note that Eq.~\eqref{local_approx} is also fully consistent with the Born-Markov approximation, which requires that the Fourier transforms of $\Gamma_\pm(\omega)$ decay quickly in time, thus they should vary slowly in frequency space on the scale of the system decoherence and thermalisation rates $\sim \alpha \omega_d$. 

\subsection{Derivation of Eq.~\eqref{eq:tilt-H-liouv}}
To derive the GME in Eq.~\eqref{eq:tilt-H-liouv} that describes the counting of excitations into the original bath, we follow similar steps to those above. We start from the counting-field dependent state of the total system in the rotating frame, which obeys the dynamical equation~\cite{Esposito2009}
\begin{equation}
    \label{cf_rho_total}
    \frac{d\hat{\rho}_{\rm tot}(\chi,t)}{dt} = -i [\hat{H}^{\prime}_{{\rm tot},\chi}, \hat{\rho}_{\rm tot}]_\chi ,
\end{equation}
where $[A_\chi, B]_\chi = A_\chi B - B A_{-\chi}$ and  $\hat{H}^{\prime}_{{\rm tot},\chi}$ is defined in Eq.~\eqref{eq:tlt-ham-rc}. The transformation to the ``interaction picture'' in Eq.~\eqref{int_picture} is now replaced by 
\begin{equation}
\label{tilted_interaction_picture}
    \tilde{\rho}_{\rm tot}(\chi,t) = e^{i\tilde{H}_{0,\chi}t}\hat{\rho}_{\rm tot}(\chi, t)e^{-i\tilde{H}_{0,-\chi}t},
\end{equation}
where $\tilde{H}_{0,\chi} = \tilde{H}_{\rm ES,\chi} + \sum_k \delta_k \hat{b}^\dagger_k \hat{b}_k$ and $\tilde{H}_{\rm ES,\chi} $ is defined below Eq.~\eqref{eq:tilt-H-liouv}. Under the Born-Markov approximation and second-order perturbation theory in $\hat{V}$, we obtain the analogue of Eq.~\eqref{BM_master_equation},
\begin{equation}
\label{tilted_BME}
      \frac{d\tilde{\rho}}{dt} = -\int_0^\infty d\tau\, {\rm tr}_{\rm B^{\prime}}[\tilde{V}_\chi(t), [\tilde{V}_\chi(t-\tau), \tilde{\rho}(t) \hat{\rho}_{\rm B^{\prime}} ]_\chi]_\chi,
\end{equation}
where $\tilde{V}_\chi(t) = e^{i\tilde{H}_{0,\chi}t}\hat{V}e^{-i\tilde{H}_{0,\chi}t} $ and similarly for $\tilde{V}_{-\chi}(t)$ (note this is a unitary transformation, unlike Eq.~\eqref{tilted_interaction_picture}). Expanding the above equation as in Eq.~\eqref{me_eigenops} is more tedious due to the counting-field dependence; schematically, we obtain terms like
\begin{align}
    \frac{d\tilde{\rho}}{dt} =  \sum_{\omega,\omega^{\prime}} & \left(\Gamma_-(\omega) e^{i(\omega^{\prime}-\omega)t}\left[\hat{a}_\chi(\omega)\tilde{\rho}\hat{a}_{-\chi}^\dagger(\omega^{\prime}) \right. \right. \notag \\  \qquad - & \left.\left.\hat{a}_\chi^\dagger(\omega^{\prime}) \hat{a}_\chi(\omega)\tilde{\rho}\right] \right) + (\rm similar \, terms),
\end{align}
with counting-field dependent ladder operators obeying
\begin{equation}
    \label{tilted_ladder_ops}
    [\tilde{H}_{\rm ES,\chi}, \hat{a}_\chi(\omega)] = -\omega \hat{a}_\chi(\omega).
\end{equation}
The transition frequencies $\omega$ are independent of $\chi$ because $\tilde{H}_{\rm ES,\chi}$ and $\tilde{H}_{\rm ES}$ are unitarily related and have the same eigenvalues. Therefore, under the same assumptions that $\Delta_{q,c}, \lambda, \Omega \ll \omega_d, T$ and $T\lesssim \omega_d$, we make the approximation in Eq.~\eqref{local_approx}. Eq.~\eqref{eq:tilt-H-liouv} is then recovered after returning to the original frame by inverting the transformation~\eqref{tilted_interaction_picture} and using $\sum_\omega \hat{a}_\chi(\omega) = \hat{a}$.

\subsection{Derivation of Eq.~\eqref{eq:tlt-D-liouv}}

To obtain the GME for counting excitation transfer to the residual bath, we again follow similar steps to those above. The counting-field dependent state now obeys the evolution equation (in the rotating frame) \begin{equation}
    \label{cf_rho_residual}
    \frac{d\hat{\rho}_{\rm tot}(\chi^{\prime},t)}{dt} = -i [\hat{H}^{\prime}_{{\rm tot},\chi^{\prime}}, \hat{\rho}_{\rm tot}]_{\chi^{\prime}} ,
\end{equation}
where $\hat{H}^{\prime}_{{\rm tot},\chi^{\prime}}$ is defined in Eq.~\eqref{eq:tlt-ham-rc-res-bath} and $\chi^{\prime}$ counts excitations transferred between the residual bath $\rm B^{\prime}$ and the extended system. Now the counting field dependence appears only in the interaction between the RC and the residual bath: $\hat{V}_{\chi^{\prime}} = \sum_k f_k e^{i\chi^{\prime}/2} \hat{a} \hat{b}_k^\dagger + \rm h.c.$ We move to the interaction picture via the transformation in Eq.~\eqref{int_picture}. Using second-order perturbation theory under the Born-Markov approximation, we obtain the analogue of Eqs.~\eqref{BM_master_equation} and~\eqref{tilted_BME}, 
\begin{equation}
\label{tilted_BME2}
      \frac{d\tilde{\rho}}{dt} = -\int_0^\infty d\tau\, {\rm tr}_{\rm B^{\prime}}[\tilde{V}_{\chi^{\prime}}(t), [\tilde{V}_{\chi^{\prime}}(t-\tau), \tilde{\rho}(t) \hat{\rho}_{\rm B^{\prime}} ]_{\chi^{\prime}}]_{\chi^{\prime}},
\end{equation}
where $\tilde{V}_{\chi^{\prime}}(t) = e^{i\tilde{H}_0t}\hat{V}_{\chi^{\prime}}e^{-i\tilde{H}_0t}$  (cf. Eq.~\eqref{int_picture}). Expanding in ladder operators of $\tilde{H}_{\rm ES}$, we obtain
\begin{widetext}
\begin{align}
\label{tilted_me_eigenops}
    \frac{d\tilde{\rho}}{dt} & = \sum_{\omega,\omega^{\prime}} \left(\Gamma_-(\omega) e^{i(\omega^{\prime}-\omega)t}\left[ 2e^{i\chi^{\prime}}\hat{a}(\omega)\tilde{\rho}\hat{a}^\dagger(\omega^{\prime}) - \left\{\hat{a}^\dagger(\omega^{\prime}) \hat{a}(\omega),\tilde{\rho}\right\}\right] + 
    \Gamma_+(\omega) e^{i(\omega^{\prime}-\omega)t}\left[2 e^{-i\chi^{\prime}}\hat{a}^\dagger(\omega)\tilde{\rho}\hat{a}(\omega^{\prime}) -\left\{\hat{a}^\dagger(\omega^{\prime}) \hat{a}(\omega),\tilde{\rho}\right\}\right]\right),
\end{align}
\end{widetext}
where all symbols have the same meaning as in Eq.~\eqref{me_eigenops} and we have already neglected the small imaginary parts of $\Gamma_\pm(\omega)$. The approximation~\eqref{local_approx} is therefore justified under the same conditions as in Appendix~\ref{me_derivation}, and so we recover Eq.~\eqref{eq:tlt-D-liouv} after moving back from the interaction picture. 
}

\renewcommand{\theequation}{B\arabic{equation}}
\renewcommand{\thefigure}{B\arabic{figure}}

\section{Detailed proof for FCS equivalence}
\label{ap:FCS_equivalence}
As discussed in Sec.~\ref{sec:fcs-original-bath}, Eq.~\eqref{eq:tilt-H-liouv} corresponds to the tilted Liouvillian $\mathcal{L}_\chi$ with counting field $\chi$ that accounts for the statistics of the net excitation number $N_{\rm B}$ exchanged between the system and the original bath via RC mode. Furthermore, in Sec.~\ref{sec:fcs-residual-bath}, Eq.~\eqref{eq:tlt-D-liouv} gives the tilted Liouvillian $\mathcal{L}^{\prime}_{\chi^{\prime}}$ with a different counting field $\chi^{\prime}$, that monitors the net excitation number $N_{\rm B^{\prime}}$ exchanged between the extended system and the residual bath. Here, we give a detailed proof for the result stated in Sec.~\ref{sec:fcs-equivalence} that, in the steady state, the particle exchange statistics is identical between the qubit-RC and the RC-residual bath. In order to achieve this, it is enough to show that the tilted Liouvillian $\mathcal{L}_\chi$ and $\mathcal{L}^{\prime}_{\chi^{\prime}}$ are related by a unitary transformation, as stated in Eq.~\eqref{eq:FCS-equivalence}.

We start from the tilted Liouvillian in Eq.~\eqref{eq:tilt-H-liouv},
\begin{equation}
    \mathcal{L}_\chi\hat{\rho}= -i[\tilde{H}_{\rm ES, \chi},\hat{\rho}]_\chi +\gamma(n_{\rm B}+1) \mathcal{D}[\hat{a}] \hat{\rho} + \gamma(n_{\rm B}) \mathcal{D}[\hat{a}^\dagger]\hat{\rho},
\end{equation}
where, $[\tilde{H}_{\rm ES, \chi},\hat{\rho}]_\chi = \tilde{H}_{\rm ES, \chi} \hat{\rho} - \hat{\rho}\tilde{H}_{\rm ES,-\chi}$ and
\begin{align}
    \tilde{H}_{\rm ES, \chi} =&~ \Delta_q\hat{\sigma}_+\hat{\sigma}_- + \Omega (\hat{\sigma}_+ +\hat{\sigma}_-) + \Delta_{c} \hat{a}^\dagger \hat{a}  \notag\\
    &+ \lambda(e^{i\chi/2}~\hat{a}^\dagger\hat{\sigma}_- +e^{-i\chi/2} ~\hat{\sigma}_+\hat{a}).
\end{align}
We then follow the arguments in Ref.~\cite{Kalaee2021}, beginning with the definition of the following unitary superoperators,
\begin{align}
    \label{eq:unitary-sup-1}
    & \mathcal{U}\hat{\rho} = e^{-\frac{i}{2} \chi \hat{a}^\dagger \hat{a}} \hat{\rho} e^{-\frac{i}{2} \chi \hat{a}^\dagger \hat{a}}, \notag \\ 
    & \mathcal{U}^\dagger\hat{\rho} = e^{\frac{i}{2} \chi \hat{a}^\dagger \hat{a}} \hat{\rho} e^{\frac{i}{2} \chi \hat{a}^\dagger \hat{a}}.
\end{align}
To transform the Liouvillian under this unitary, we will now work with the following vectorised forms of the super operators,
\begin{align}
    \mathcal{L}_{\chi}  &= -i (\mathds{1} \otimes \tilde{H}_{\rm ES, \chi} -  \tilde{H}_{\rm ES, -\chi}^{\rm{T}}  \otimes \mathds{1} ) \notag \\
    & ~~~~ + \gamma_1 ({\hat{a} \otimes \hat{a}} - \frac{1}{2} \mathds{1} \otimes \hat{a}^\dagger \hat{a} - \frac{1}{2}  (\hat{a}^\dagger \hat{a})^{\rm{T}} \otimes \mathds{1} ) \notag \\
    &~~~~ + \gamma_2 ({\hat{a} \otimes \hat{a}} - \frac{1}{2} \mathds{1} \otimes \hat{a} \hat{a}^\dagger - \frac{1}{2}  (\hat{a} \hat{a}^\dagger)^{\rm{T}} \otimes \mathds{1} ) \notag \\
    &=  -i (\mathds{1} \otimes \tilde{H}_{\rm ES, \chi} -  \tilde{H}_{\rm ES, \chi}  \otimes \mathds{1} ) \notag \\
    &~~~~ + \gamma_1 ({\hat{a} \otimes \hat{a}} - \frac{1}{2} \mathds{1} \otimes \hat{a}^\dagger \hat{a} - \frac{1}{2}  \hat{a}^\dagger \hat{a} \otimes \mathds{1} ) \notag \\
    &~~~~ + \gamma_2 ({\hat{a}^\dagger \otimes \hat{a}^\dagger} - \frac{1}{2} \mathds{1} \otimes \hat{a} \hat{a}^\dagger - \frac{1}{2}  \hat{a} \hat{a}^\dagger \otimes \mathds{1} ).
\end{align}
where, $\gamma_1 = \gamma(n_{\rm B} +1)$ and $\gamma_2 = \gamma(n_{\rm B})$. We identify the unitary operator $U = e^{-\frac{i}{2}\chi \hat{a}^\dagger \hat{a}}$, so the vectorised form of the unitary superoperator $\mathcal{U}$ becomes,
\begin{align}
    & \mathcal{U} = U\otimes U =  e^{-\frac{i}{2} \chi \hat{a}^\dagger \hat{a}} \otimes e^{-\frac{i}{2} \chi \hat{a}^\dagger \hat{a}}, \notag \\ 
    & \mathcal{U}^\dagger =U^\dagger \otimes U^\dagger = e^{\frac{i}{2} \chi \hat{a}^\dagger \hat{a}} \otimes e^{\frac{i}{2} \chi \hat{a}^\dagger \hat{a}}.
\end{align}
The unitary frame transforms the Liouvillian as $ \mathcal{L}_\chi \to\mathcal{U}^\dagger \mathcal{L}_\chi \mathcal{U}$. It is straightforward to calculate this by noting the following relations,
\begin{align}
    & \hat{U}^\dagger \tilde{H}_{\rm ES,\chi} \hat{U} = \tilde{H}_{ES}, \quad \hat{U}^\dagger \hat{a}^\dagger \hat{a} \hat{U}  =  \hat{a}^\dagger \hat{a}, \notag \\
    &\hat{U}^\dagger  \hat{a}  \hat{U} =  e^{-\frac{i}{2}\chi}\hat{a},  \quad
    \hat{U}^\dagger   \hat{a}^\dagger \hat{U} = e^{\frac{i}{2}\chi}\hat{a}^\dagger .
\end{align}
Finally the transformed Liouvillian in the vectorised form reads,
\begin{align}
    \mathcal{U}^\dagger \mathcal{L}_\chi \mathcal{U} &=  -i (\mathds{1} \otimes \tilde{H}_{ES} -  \tilde{H}_{ES}  \otimes \mathds{1} ) \notag \\
    &~~~~ + \gamma_1 ({e^{\frac{i}{2}\chi}\hat{a} \otimes e^{\frac{i}{2}\chi} \hat{a}} - \frac{1}{2} \mathds{1} \otimes \hat{a}^\dagger \hat{a} - \frac{1}{2}  \hat{a}^\dagger \hat{a} \otimes \mathds{1} ) \notag \\
    &~~~~ + \gamma_2 ({e^{-\frac{i}{2}\chi}\hat{a}^\dagger \otimes e^{-\frac{i}{2}\chi} \hat{a}^\dagger} - \frac{1}{2} \mathds{1} \otimes \hat{a} \hat{a}^\dagger - \frac{1}{2}  \hat{a} \hat{a}^\dagger \otimes \mathds{1} ).
\end{align}
By noticing that the RHS in the above equation is the vectorised form of the tilted Liouvillian $\mathcal{L}^{\prime}_{\chi^{\prime}}$ with $\chi^{\prime}=\chi$, we finally recover the relation in Eq.~\eqref{eq:FCS-equivalence},
\begin{align}
    \mathcal{U}^\dagger\mathcal{L}_\chi \mathcal{U} &= -i[\tilde{H}_{\rm ES}, \hat{\rho}] 
    + \gamma (n_{\rm B} +1 )\mathcal{D}_{\chi}[\hat{a}] \hat{\rho} + \gamma (n_{\rm B})\mathcal{D}_{-\chi}[{\hat{a}^\dagger}] \hat{\rho}\notag \\
    &= \mathcal{L}^{\prime}_{\chi}.
\end{align}

\renewcommand{\theequation}{C\arabic{equation}}
\renewcommand{\thefigure}{C\arabic{figure}}

{
\section{Thermodynamic Hamiltonian}
\label{ap:thermodynamic_hamiltonian}
In this appendix, we provide further justification for the form of the thermodynamic Hamiltonian~\eqref{thermodynamic_Hamiltonian} and the work and heat defined in Eqs.~\eqref{heat} and \eqref{work}. We start by examining the power as defined by the standard relation~\cite{alickiQuantumOpenSystem1979}
\begin{equation}
    \label{alicki_power}
    \dot{W} = \left\langle \frac{\partial \hat{H}}{\partial t}\right \rangle. 
\end{equation}
In the lab frame, with Hamiltonian given by Eq.~\eqref{eq:driven_qubit_total}, we have
\begin{equation}
    \label{lab_power}
    \dot{W} = -2\omega_d\Omega \sin(\omega_d t)\langle \hat{\sigma}_x\rangle.
\end{equation}
The expectation value is invariant under a unitary transformation into the rotating frame, $\hat{U}(t) = \exp(i\omega_d\hat{N}_{\rm tot}t)$, defined above Eq.~\eqref{eq:RC_driven_qubit_rot}. We therefore find 
\begin{equation}
    \label{rotating_frame_power}
    \dot{W} = -2\omega_d\Omega \sin(\omega_d t) {\rm tr}\left( e^{i\omega_d t}\hat{\sigma}_+\hat{\rho}\right) + \rm c.c.,
\end{equation}
where $\hat{\rho}$ is the density matrix in the rotating frame. Within the RWA, the fast-oscillating counter-rotating terms at frequency $\pm 2\omega_d$ are neglected, leaving 
\begin{equation}
    \label{power_RWA}
    \dot{W} = -i\omega_d \Omega \left(\hat{\sigma}_+\hat{\rho}\right) + {\rm c.c.} = \omega_d \Omega \langle \hat{\sigma}_y\rangle.
\end{equation}
A direct calculation shows that this is equivalent to Eq.~\eqref{work}. Therefore, we can consistently identify the internal energy $U = {\rm tr}[\hat{H}_{\rm TD} \hat{\rho}]$ with the expectation value of the thermodynamic Hamiltonian~\eqref{thermodynamic_Hamiltonian}, so long as we define the heat flux as in Eq.~\eqref{heat}, i.e.
\begin{equation}
\label{ap_heat}
    \dot{Q} = \dot{U} - \dot{W} = {\rm tr}[\hat{H}_{\rm TD} \mathcal{D}\hat{\rho}].
\end{equation}

To show that the heat flux~\eqref{ap_heat} is proportional to the excitation current, we use the master equation~\eqref{eq:RC-LME-driv-qbit} to directly compute
\begin{equation}
\label{heat_diss}
    \dot{Q} = {\rm tr}[\hat{H}_{\rm TD} \mathcal{D}\hat{\rho}(t)] = \gamma \omega_d \left( n_{\rm B} -{\rm tr}\left[\hat{a}^\dagger \hat{a}\hat{\rho}(t)\right] \right).
\end{equation}
The mean excitation current into the residual bath is given within the FCS formalism (see Sec.~\ref{sec:fcs}) by 
\begin{align}
    \label{mean_current_FCS}
   J(t) =  \frac{d}{dt}\mathrm{E}[N_{\rm B^{\prime}}(t)] & = -i \left.\frac{d}{d\chi^{\prime}}\right\rvert_{\chi^{\prime}=0}{\rm tr}\left[\mathcal{L}^{\prime}_{\chi^{\prime}} \hat{\rho}(\chi^{\prime},t) \right] \notag \\
    & = \gamma  \left({\rm tr}\left[\hat{a}^\dagger \hat{a}\hat{\rho}(t)\right] - n_{\rm B} \right),
\end{align}
where the final equality follows from the explicit expression for $\mathcal{L}^{\prime}_{\chi^{\prime}}$ given in Eq.~\eqref{eq:tlt-dissipator}. Comparing Eqs.~\eqref{heat_diss} and~\eqref{mean_current_FCS}, we conclude that $\dot{Q} =-\omega_d J(t)$. 

This proportionality between the heat and excitation currents is fully consistent with the assumptions underpinning the master equations derived in Appendix~\ref{ap:me_derivation_sketch}. In particular, we assume that the bath spectral functions are approximately constant on the scale of the decay rates $\sim \gamma$ {and} also the small frequency splittings $\omega \sim \lambda,\Omega,\Delta_{q,c}$ around the central drive frequency $\omega_d$. Therefore, the incoherent dynamics does not resolve these small energy differences and we can consistently assign an energy $\omega_d$ to each excitation transferred to the residual bath.

}

\bibliography{biblio}

\end{document}